\documentclass[aps,prl,twocolumn,showpacs,10pt,superscriptaddress,groupedaddress, footinbib]{revtex4-1}

\usepackage{amsmath,bm,amssymb}
\usepackage{amsbsy}
\usepackage{autobreak}
\usepackage{endnotes}
\let\footnote=\endnote

\usepackage{graphicx}
\usepackage{psfrag}
\usepackage[caption=false]{subfig}
\usepackage{color}
\definecolor{urlblue}{rgb}{0.2,0.4,0.7}
\definecolor{citegreen}{rgb}{0,0.6,0.2}
\usepackage[caption=false]{subfig}
\usepackage{color}
\definecolor{urlblue}{rgb}{0.2,0.4,0.7}
\definecolor{citegreen}{rgb}{0,0.6,0.2}
\definecolor{linkred}{rgb}{0.9,0.2,0.1}
\usepackage{hyperref}
\hypersetup{
colorlinks=true, citecolor=citegreen, linkcolor=blue,urlcolor=urlblue}
\usepackage[table]{xcolor}
\usepackage{tabularx}
\newcolumntype{P}[1]{>{\centering\arraybackslash}p{#1}}
\usepackage{multirow, caption, booktabs}
\usepackage{slashed}
\usepackage{array}
\newcolumntype{P}[1]{>{\centering\arraybackslash}p{#1}}
\allowdisplaybreaks

\def\zo{\overline{z}_1}
\def\zt{\overline{z}_2}

\definecolor{linkred}{rgb}{0.9,0.2,0.1}
\usepackage{hyperref}
\hypersetup{
colorlinks=true, citecolor=citegreen, linkcolor=blue,urlcolor=urlblue}

\usepackage[table]{xcolor}
\usepackage{tabularx}
\newcolumntype{P}[1]{>{\centering\arraybackslash}p{#1}}
\usepackage{multirow, caption, booktabs}

\usepackage{slashed}
\usepackage{array}
\newcolumntype{P}[1]{>{\centering\arraybackslash}p{#1}}

\allowdisplaybreaks

\def\zo{\overline{z}_1}
\def\zt{\overline{z}_2}

\begin{document}


\begin{flushleft} 
{ \mbox{IMSc/2022/03} }
\end{flushleft}

\title{Resummed next-to-soft corrections to rapidity distribution of Higgs Boson to $ \textbf{NNLO} + \overline{\textbf{NNLL} }  $ }

\author{V. Ravindran\footnote{ravindra@imsc.res.in}}
\affiliation{The Institute of Mathematical Sciences, HBNI, Taramani,
 Chennai 600113, India}
\author{Aparna Sankar\footnote{aparnas@imsc.res.in}}
\affiliation{The Institute of Mathematical Sciences, HBNI, Taramani,
 Chennai 600113, India}
\author{Surabhi Tiwari\footnote{surabhit@imsc.res.in}}
\affiliation{The Institute of Mathematical Sciences, HBNI, Taramani,
 Chennai 600113, India}

\date{\today}

\begin{abstract}
We present the resumed predictions consisting of both soft-virtual(SV) as well as next-to-SV(NSV) threshold logarithms to all orders in perturbative QCD for the rapidity distribution of Higgs Boson till $\rm NNLO + \overline{NNLL}$ accuracy at LHC. Using our recent formalism\cite{Ajjath:2020lwb}, the resummation is carried out in the double Mellin space by restricting the NSV contributions only from diagonal $gg$ channel. We perform the inverse Mellin ransformation using the minimal prescription procedure and match it with the corresponding fixed order results. We do a detailed analysis of the numerical impact of the resummed result. The K-factor values at different logarithmic accuracy suggest that the prediction for the rapidity distribution converges and becomes more reliable at $\rm NNLO + \overline{NNLL}$ order. We further observed that the inclusion of resumed NSV contribution improves the renormalisation scale uncertainty at every order in perturbation theory. However, the uncertainty due to factorisation scale increases by the addition of resummed SV+NSV predictions to the fixed order rapidity distribution.
\end{abstract}

\maketitle
\section{Introduction}
There have been plethora of data available on accurate measurements of observables from LHC at CERN in recent times. This combined with the precise theoretical predictions from various state-of-the-art computations have facilitated in establishing Standard Model(SM) as being extremely successful in describing the physics of elementary particles. It has also helped in probing physics beyond the SM(BSM) scenarios in a very clear environment. In the next few years, the High-Luminosity LHC will come into effect which will not only increase the chances to see rare processes but will also improve the precision of measurements. One of the most important process at LHC is the Higgs boson production which helps in probing the electroweak symmetry breaking and the coupling of the Higgs boson with other SM particles. The dominant channel in Higgs production is the gluon fusion process due to the large flux of gluons present in the protons at these energies. The other alternate channel is through the bottom quark annihilation which has gained attention of theoretical physicists in recent times due to the freedom it provides in treating the initial state bottom-quarks. The Drell-Yan production of a pair of leptons through the decay of virtual photons, Z and W bosons at LHC is also an effective processe as it helps in probing the structure of hadrons. 

The measurements of inclusive and differential rates \cite{Aad:2012tfa,Chatrchyan:2012xdj} like transverse momentum and rapidity distributions of the Higgs boson production are very useful in understanding the symmetry breaking mechanism and Higgs boson coupling with other SM particles. Likewise, measurements \cite{Affolder:2000rx,Abe:1998rv,CMS:2014jea} pertaining to Drell-Yan production for inclusive as well as differetial rates provide pivotal information related to BSM scenarios namely, R-parity violating supersymmetric models \cite{CDF:2001gjd}, models with $Z'$ or with contact interactions, and large extra-dimension models \cite{ArkaniHamed:1998rs,Randall:1999ee}. The rapidity distributions of Drell-Yan production have also been used since a long time, to calibrate the detectors and to obtain the parton distribution functions(PDFs) \cite{Gao:2013xoa,Harland-Lang:2014zoa,Ball:2014uwa,Butterworth:2015oua,Alekhin:2017kpj}. The second order(NNLO) QCD predictions for inclusive and rapidity distribution for both Drell-Yan and Higgs Boson production in gluon-gluon fusion has been known for a long time. In recent years, the third order(N3LO) result for Drell-Yan \cite{Duhr:2020seh,Ahmed:2014cla} as well as Higgs production through gluon fusion \cite{Anastasiou:2015vya,Li:2014bfa,Catani:2014uta} and through bottom quark annihilation \cite{Duhr:2019kwi,Ahmed:2014era}, have become available for both these observables.

The fixed order QCD predictions for both inclusive and differential cross-sections have limitations in applicability due to the presence of various logarithms which become large in certain kinematical regions called the threshold region. This threshold limit results from the emission of the soft gluons in the case of inclusive cross-section for DY and Higgs Boson production. Whereas, for the transverse momentum computation, it arises when the final state becomes small. These dominant contributions in the form of logarithms spoil the reliability of the perturbative results from the truncated series. A viable solution to this problem is to systematically resume these large logarithms to all orders in perturbation theory and then supplementing it with the fixed order results can cover the entire kinematic region of the phase space. There are several approaches in literature to achieve this for both inclusive as well as differential distributions \cite{Sterman:1986aj,Catani:1989ne,Catani:1990rp,Moch:2005ky,Laenen:2005uz,Idilbi:2005ni,Ravindran:2005vv}. Sterman \cite{Sterman:1986aj} and, Catani and Trentadue \cite{Catani:1989ne} did extensive studies in their seminal work to achieve the resummation of these leading large logarithms also called soft-virtual(SV) logarithms through reorganisation of the perturbative series. They performed this threshold resummation in Mellin space for the inclusive case whereas for the rapidity distribution, the formalism was extended using double Mellin moments. Using factorization properties and renormalisation group(RG) invariance, one of the authors of the present paper developed an all order $z$-space formalism to capture the threshold-enhanced contribution to inclusive production \cite{Ravindran:2005vv,Ravindran:2006cg} as well as rapidity distribution \cite{Ravindran:2006bu} of any colorless particle. This formalism was further used for Z and $\rm W^{\pm}$ case \cite{Ravindran:2007sv} using two scaling variables. It was also applied to Drell Yan and Higgs Boson production at $\rm N^3LO$ level \cite{Ahmed:2014uya,Ahmed:2014era} and at NNLO + NNLL accuracy \cite{Banerjee:2017cfc,Banerjee:2018vvb}. The threshold resummation technique is developed using soft-collinear effective theory(SCET) as well. Here, the resummation is performed in momentum space for inclusive \cite{Idilbi:2005ky} as well as transverse momentum distribution \cite{Becher:2010tm}. The resummation for rapidity distribution using SCET formalism has been carried out in Refs. \cite{Becher:2006nr, Becher:2007ty, Bonvini:2014qga} .

The threshold resummation technique that we have followed in this paper is based on Refs. \cite{Ravindran:2006bu, Ravindran:2007sv, Catani:1989ne}. Here, the resummation for the rapidity distribution is done in the two-dimensional Melin space as the convolutions become normal products in the $N$-space. The double Mellin variables $N_1$ and $N_2$ correspond to $z_1$ and $z_2$ in $z$ space and hence the resummation of large logarithms becomes proportional to $\log(N_i)$ in the limit $N_i \rightarrow \infty \, (z_i \rightarrow 1)$. In the large N limit, the $\log(N_i)$ combined with the strong coupling constant $a_s$ gives order one terms and therefore truncating the series at a particular order of $a_s$ is not possible. This difficulty is overcome by using the factorisation properties, universality of IR contributions and renormalisation group invariance to systematically resum these order one terms to all orders in perturbation theory. Here, we not only deal with the SV distributions but have included the resummed subleading threshold logarithms known as next to soft-virtual(NSV) logarithms as well. The importance of these collinear NSV logarithms was understood long ago and several attempts have been made so far to understand the structure of these corrections for certain inclusive\cite{Anastasiou:2015ema,Anastasiou:2015vya,Duhr:2019kwi,Duhr:2020seh} and differential\cite{Dulat:2018bfe} observables. Unlike SV distributions, these logarithms are also present in the off-diagonal channels. The understanding of their all-order structure is still an open problem\cite{Laenen:2008ux,Laenen:2010uz,Bonocore:2014wua,Bonocore:2015esa,Bonocore:2016awd,DelDuca:2017twk,Bahjat-Abbas:2019fqa,Soar:2009yh,Moch:2009hr,deFlorian:2014vta,Beneke:2018gvs,Beneke:2019mua,Beneke:2019oqx}. Recently, we developed a formalism to systematically resum the NSV logarithms coming from diagonal channels for inclusive cross-section of various processes at LHC\cite{Ajjath:2020ulr,Ajjath:2020sjk,Ajjath:2021bbm,Ajjath:2021lvg}. It was later extended for the case of rapidity distribution of a pair of leptons in Drell-Yan and a Higgs boson in gluon fusion as well as in bottom quark annihilation as well\cite{Ajjath:2020lwb}. We achieved the resummation of these NSV logarithms to all orders in perturbation theory in z as well as in the Mellin N space.

In this paper, we present the phenomenological importance of the resummed NSV logarithms for the production of Higgs Boson via gluon fusion at LHC. We first give the theoretical overview of the formalism developed to study these logarithms along with the relevant theoretical results. This is followed by the comprehensive study on the numerical impact of NSV contributions at various orders in perturbation theory. Finally, we conclude the paper by providing a short discussion on our main findings.

\section{Theoretical Framework}
We begin by considering a generic hadronic collision between two hadrons $H_{1,(2)}$ having momentum $P_{1,(2)}$ that produces a colourless final state denoted as $F(q)$ with $q$ being its momentum.
\begin{align}
\label{eq:had-collision}
    H_1(P_1) + H_2(P_2) \rightarrow F(q) + X\,.
\end{align}
where the quantity $X$ represents an inclusive hadronic state. The rapidity of this final colourless state $F$, is defined through,
\begin{align}
\label{eq:rapidity}
    y \equiv \frac{1}{2} \ln \left( \frac{P_2 \cdot q}{P_1 \cdot q}\right)\,.
\end{align}

In QCD improved parton model, the differential distribution with respect to rapidity for the colorless state $F$ at a hadronic level for Higgs Boson production through gluon fusion can be expressed as,
\begin{eqnarray}\label{sighad}
{d \sigma^g\over dy } &=&
\sigma^g_{\rm B}(\tau, q^2) 
\sum_{a,b=q,\overline q,g}
\int_{x_1^0}^1 {dz_1 \over z_1}\int_{x_2^0}^1 {dz_2 \over z_2}~ 
f_{a}\left({x_1^0 \over z_1},\mu_F^2\right)
\nonumber \\ 
&& \times f_{b}\left({x_2^0\over z_2}, \mu_F^2\right)
\Delta^g_{d,ab} (z_1,z_2,q^2,\mu_F^2,\mu_R^2).
\end{eqnarray}
where $\sigma_B$ is the leading order contribution. The dimensionless variable $z$ is defined as $z = \frac{q^2}{\hat{s}}$, where $\hat{s}$ is the square of partonic centre of mass energy and $\tau$ is defined as, $\tau = q^2/S$ where $S$ is the square of hadronic centre of mass energy. The non-perturbative functions $f_c(x_l,\mu_F^2)$ are the parton distribution functions(PDFs) of the colliding partons $a,b$ with momentum fractions $x_l~(l=1,2)$. These PDFs are renormalized at the factorization scale $\mu_F$. The other scale appearing in the above equation is the renormalisation scale $\mu_R$ used for the evolution of strong couling constant $\alpha_s$. Here, we limit our computation to the region of threshold limit or soft limit i.e. $z \rightarrow 1$ and the corresponding leading contribution to the differential rapidity distribution is referred to as Soft-Virtual(SV) + Next-to-Soft-Virtual(NSV) contributions. In order to define the threshold limit at the partonic level for the rapidity distribution, we choose to work with a set of symmetric scaling variables $x^{0}_{1(2)}$ instead of $y$ and $\tau$ which are related through,
\begin{align}
\label{eq:tau-y-x0}
    y \equiv \frac{1}{2} \ln\left( \frac{x_1^0}{x_2^0}\right) \quad \text{and} \quad \tau \equiv x_1^0 x_2^0\,.
\end{align}

In \eqref{sighad}, $\Delta^g_{d,ab}$ are calculated perturbatively in powers of strong coupling constant $a_s = g^2_s/16\pi^2$ and are referred to as coefficient functions(CF). Beyond leading order, the CFs contain ultraviolet(UV) and infrared(IR) divergences at the intermediate stages. The UV divergences are often removed in the $\overline{MS}$ renormalisation scheme at a renormalisation scale $\mu_R$. The UV finite partonic cross-sections are then left with two categories of IR divergences namely soft and collinear divergences. The soft divergences associated with the soft gluons get canceled between the virtual and real emission diagrams in infrared safe observables. The collinear singularities related to the collinear partons are removed by summing over degenerate final states and by mass factorization in $\overline{MS}$ scheme at a factorisation scale $\mu_F$. The fixed order perturbative results are therefore, often sensitive to scales, $\mu_R$ and $\mu_F$. The CFs contain various logarithmic structures and are represented as,
\begin{eqnarray}
\Delta_{d,ab}^g (z_1,z_2,q^2,\mu_F^2,\mu_R^2) &=&
\nonumber\\ & \hspace{-4em}
\sum_{i=0}^\infty a_s^i(\mu_R^2) \Delta^{(i),g}_{d,ab}(z_1, z_2, q^2, \mu_F^2,\mu_R^2) \,,
\end{eqnarray}
\text{where}

\begin{eqnarray}
\Delta^{(i),g}_{d,ab}(z_1, z_2, q^2, \mu_F^2,\mu_R^2) &=
\Delta^{(i),g}_{d,ab,\delta \delta} \delta(1 - z_1) \delta(1 - z_2)
\nonumber\\ & \hspace{-10em}
+ \sum \Delta^{(i),g}_{d,ab,\delta {\cal D}_j} \delta(1 - z_2) {\cal D}_j
+ \sum \Delta^{(i),g}_{d,ab,\delta \overline{{\cal D}}_j}
 \delta(1 - z_1) \overline{{\cal D}}_j
\nonumber\\ & \hspace{-10em}
+ \sum \Delta^{(i),g}_{d,ab,{\cal D}_j \overline{{\cal D}}_k}
 {\cal D}_j  \overline{{\cal D}}_k 
+ \Delta^{(i),g}_{d,ab,R}(z_1,z_2) \, , \label{delsv}
\end{eqnarray}

\begin{align}
\text{with}~
{\cal D}_i=\Bigg[{\ln^i(1-z_1) \over (1-z_1)}\Bigg]_+ \,,
\bar{{\cal D}}_i=\Bigg[{\ln^i(1-z_2) \over (1-z_2)}\Bigg]_+  .
\end{align}

The distributions ${\cal D}_i$ and $\bar{{\cal D}}_i$ along with the $\delta(1 - z_1)$ and $\delta(1 - z_2)$ part are called the soft-virtual(SV) contributions to the infrared safe observable. These logarithms come from only diagonal channels. The leading contributions of the regular part $\Delta^{(i),g}_{d,ab,R}(z_1,z_2)$  near the threshold region $z_l = 1$, consist of terms of the form ${\cal D}_i(z_l)\ln^k(1-z_j)$ and $\delta(1-z_l) \ln^k(1-z_j)$ with ($l,j=1,2),~(i,k=0,1,\cdots$). These are called next to soft-virtual(NSV) contributions. The NSV terms come from diagonal as well as non-diagonal channels. The SV terms have been studied in detail by one of the authors of this paper \cite{Ravindran:2006bu}. In the next section, we explore the NSV contributions to the partonic CFs.

\subsection{Next to SV in $z$ space}
The subleading NSV contributions present in the regular part of the partonic CFs play an important role in the precise prediction of the differential distributions. They also help in understanding the structure of beyond SV terms in the threshold expansion at higher orders. The regular part of the CFs containing beyond SV terms, is expanded around $z=1$ in the following way,
\begin{eqnarray}
\Delta^{(i),g}_{d,ab,R}(z_1, z_2) =& \prod_{j=1,2}\sum_{k=0}^{2i-1}\sum_{l=0}^\infty  \Delta^{reg,(i),g}_{ d,a b,l,k}  \times \\ \nonumber &(1-z_j)^l  \ln^k(1-z_j)
\label{delregexp}
\end{eqnarray}
The NSV logarithms also demonstrate perturbative behaviour like SV logarithms and therefore can be expressed in the powers of strong coupling constant $\alpha_s$,
\begin{eqnarray}\label{NSV}
\Delta_{d,ab}^{g,NSV}(z_1,z_2) = \sum_{i=0}^\infty a_s^i(\mu_R^2) \Delta_{d,ab}^{g,NSV,(i)}(z_1,z_2)
,\, 
\end{eqnarray}
In the threshold region near $z=1$, $\Delta_{d,ab}^{g,NSV,(i)}(z_1,z_2)$ is defined by setting $l=0$ in \eqref{delregexp} as,
\begin{eqnarray}
\label{DeltaR}
\Delta_{d,ab}^{g,NSV,(i)}(z_1,z_2) = {\displaystyle \prod_{j=1,2}} \sum_{k=0}^{2 i-1} \Delta_{d,ab,0,k}^{reg,(i),g} \ln^k(1-z_j) \,.
\end{eqnarray}
These NSV contributions are sometimes also called next-to-leading power(NLP) contributions. We recently, developed a formalism\cite{Ajjath:2020lwb} to study the all-order behaviour of NSV terms in rapidity distributions of any colorless particle produced in hadron colliders. Our formalism systematically include the NSV contributions coming from only the diagonal channel of any process. We determined the complete NSV contributions to third order in strong coupling constant for the rapidity distributions of Drell-Yan process and also for Higgs boson production via gluon fusion as well as bottom quark annihilation. The all order z-space result is exhibited in a compact form through an integral representation. This z-space integral representation is then used to resum order one terms in two-dimensional Mellin space to get a reliable theoretical prediction.

The formalism uses RG invariance and factorisation properties to show that the diagonal CFs containing both SV and NSV contributions exponentiate as,

\begin{align}\label{delta}
\Delta^{\rm SV+ \rm NSV}_{d,g} ={\cal C} \exp
\Big({\Psi^q_d(\mu_R^2,\mu_F^2,\zo,\zt,\epsilon)}\Big)\, \Big|_{\epsilon = 0} \,,
\end{align}

where the function $\Psi_d^g$ is given by
\begin{flalign}
\label{eq:Psi}
&\Psi_d^g\big(\mu_R^2,\mu_F^2,\zo,\zt,\varepsilon\big) = \bigg( \ln ( Z_{UV,g} (\hat{a}_s,\mu^2,\mu_R^2,\varepsilon ) )^2   \nonumber  \\ 
&+ \ln \big| \hat{F}_{g}\big(\hat{a}_s,\mu^2,-m_H^2,\varepsilon)\big|^2\bigg) \delta(\zo)\delta(\zt) \nonumber \\
&+2 \mathrm{\Phi}_d^g\big(\hat{a}_s,\mu^2,m_H^2,\zo,\zt,\varepsilon\big) \nonumber \\
&- \mathcal{C} \ln  \Gamma_{gg}\big(\hat{a}_s,\mu^2,\mu_F^2,\zo,\varepsilon\big)\delta(\zt)  \nonumber \\
&- \mathcal{C}\ln \Gamma_{gg}\big(\hat{a}_s,\mu^2,\mu_F^2,\zt,\varepsilon\big)\delta(\zo) \,.
\end{flalign}
The function $\Psi^g_d $ is computed in $4+\epsilon$ space-time dimensions in perturbative QCD and the scaling variables are shifted to $\zo = 1- z_1$ and $\zt = 1-z_2$. The symbol $\mathcal{C}$ refers to convolution and its action on any exponential of a function can be found in \cite{Ajjath:2020ulr}. One of the authors demonstrated the above equation in his paper\cite{Ravindran:2006bu} stating that $\Psi^g_d $ can be decomposed in terms of form factor $F^g$, soft distribution $\Phi^g_d$ and the diagonal Altarelli-Parisi kernels $\Gamma_{gg}$. The function $Z_{UV,g}$ is the overall renormalisation constant. The different terms in $\Psi_d^g$ are UV and IR divergent individually. However, when these terms get summed up, the divergences cancel among each other making $\Psi_d^g$ finite and regular in the variable $\epsilon$. The soft distribution function containing only the soft contribution associated with the real emission has been discussed in great detail in \cite{Ravindran:2006bu}. Using our formalism, we showed in Eq.(3) of \cite{Ajjath:2020lwb} that the K+G type Sudakov differential equation admits solution that encapsulate both soft and next-to-soft contributions. The divergent part of the NSV contribution to the soft distribution function cancels against the collinear singularities from AP kernels. Using the NSV incorporated soft distribution function $\mathrm{\Phi}_d^g$, we get the integral representation of the finite function $\Psi^g_d$ which contains the all order information of the mass-factorised differential distribution,


\begin{align}
\label{eq:psiint}
\Psi^g_d =& {\delta(\overline z_1) \over 2} \Bigg(\!\!\displaystyle {\int_{\mu_F^2}^{q^2 \overline z_2}
\!\!{d \lambda^2 \over \lambda^2}}\! {\cal P}^g\left(a_s(\lambda^2),\zt\right) 
\!+\! {\cal Q}^g_d\left(a_s(q_2^2),\zt\right)
\!\!\Bigg)_+ 
\nonumber\\&
+ {1 \over 4} \Bigg( {1 \over \overline z_1 }
\Bigg\{{\cal P}^g(a_s(q_{12}^2),\zt ) + {\color{black} 2 }L^g(a_s(q_{12}^2) ,\zt)
\nonumber \\
& + q^2{d \over dq^2} 
\left({\cal Q}^{g}_d(a_s(q_{12}^2 ),\zt) +  {\color{black} 2 }\varphi_{d,g}^f(a_s(q_{12}^2 ),\zt)\right)
\Bigg\}\Bigg)_+
\nonumber\\&
+ {1 \over 2}
\delta(\overline z_1) \delta(\overline z_2)
\ln \Big(g^g_{d,0}(a_s(\mu_F^2))\Big)
+ \overline z_1 \leftrightarrow \overline z_2,
\end{align}
where ${\cal P}^g (a_s, \overline z_l)= P^g(a_s,\overline z_l) - 2 B^g(a_s) \delta(\overline z_l)$, $q_l^2 = q^2~(1-z_l) $ and $q^2_{12}=q^2 \overline z_1 \overline z_2$. The subscript $+$ indicates standard plus distribution.
We consider only the diagonal parts of the AP splitting function $P^g(a_s,\overline z_l)$ as non-diagonal splitting functions upon convolutions give only beyond NSV terms. The diagonal AP splitting function  near $z=1$ is given as,
\begin{align}
&P_{gg}\big(z_j,a_s(\mu_F^2)\big) = 2  \bigg[ B^g(a_s(\mu_F^2)) \delta(1-z_j) + \nonumber \\ &A^g(a_s(\mu_F^2)) {\cal D}_0(z_j) + L^g(a_s(\mu_F^2),z_j) \bigg]
\end{align}
where $A^g$ and $B^g$ are the cusp and collinear anomalous dimensions respectively and,
\begin{align}
L^g(a_s(\mu_F^2), z_j) \equiv C^g(a_s(\mu_F^2)) \ln(1 - z_j)  + D^g(a_s(\mu_F^2))
\end{align}
The cusp, the collinear anomalous dimensions and  the constants $C^g$ and $D^g$ are available in \cite{Moch:2004pa,Vogt:2004mw,Blumlein:2021enk,Ajjath:2021lvg} till third order. The constant $g^g_{d,0}$ present in Eq.\eqref{eq:psiint} encapsulates finite part of the virtual contributions and pure $\delta(\overline z_l)$ terms of the soft distribution function $\Phi^g_d$.

The function ${\cal Q}^g_d$ present in Eq.\eqref{eq:psiint} is expressed as,
\begin{eqnarray}
{\cal Q}^g_d(a_s,\overline z_l) = {2 \over \overline z_l}  D_d^g(a_s) + 2 \varphi^f_{d,g} (a_s,\overline z_l)\,.
\end{eqnarray}
The functional form of SV coefficent $D_d^g$ is given in Eq.(7) of Ref.\cite{Banerjee:2017cfc} where it is expanded in powers of strong coupling constant $\alpha_s$ in the limit $\epsilon \rightarrow 0$ and presented till third order. The function $\varphi_{d,g}$ is the finite part of the NSV contribution  $\varphi_{d,g}^{(i)}$ to the soft distribution function given in Eq.(3) of \cite{Ajjath:2020lwb} and is parametrized in terms of $\ln^k(1-z_j)$ as,
\begin{align}
\label{eq:Phidf}
\varphi_{d,g}^f(a_s(\lambda^2),\overline z_l) &= \sum_{i=1}^\infty \sum_{k=0}^{\infty} \hat  a_s^i \left({\lambda^2 \over \mu^2}\right)^{i\frac{\epsilon}{2}}
S_\epsilon^i 
\varphi^{(i,k)}_{d,g}(\epsilon)\ln^k \overline z_l\,,
\nonumber\\
&= 
\sum_{i=1}^\infty \sum_{k=0}^i a_s^i(\lambda^2) \varphi^{g,(k)}_{d,i} \ln^k \overline z_l\,.
\end{align} 
The upper limit on the sum over $k$ is controlled by the dimensionally regularised Feynman integrals that contribute to order $a_s^i$. The coefficients $\varphi^{g,(k)}_{d,i}$ given in the above equation are known till third order and listed below, 
\begin{widetext}
\begin{align}
\label{eq:Phig}
\varphi^{g,(0)}_{d,1} &= 
       2 C_A  \,,
\nonumber \quad        
\varphi^{g,(1)}_{d,1} = 0\,,
\nonumber \quad  
\varphi^{g,(0)}_{d,2} =
        C_A n_f   \bigg(  - \frac{136}{27} + \frac{8}{3} \zeta_2 \bigg) + C_A^2   \bigg( \frac{904}{27} - 28 \zeta_3 - \frac{104}{3} \zeta_2 \bigg)\,,
\nonumber\\
\varphi^{g,(1)}_{d,2} &=
        C_A n_f   \bigg(  - \frac{2}{3} \bigg) + C_A^2   \bigg( \frac{2}{3} \bigg)\,,
\nonumber \quad        
\varphi^{g,(2)}_{d,2} = 
        C_A^2   \bigg(  - 4 \bigg)\,,
\nonumber\\
\varphi^{g,(0)}_{d,3} &=
        C_A n_f^2   \bigg(  - \frac{232}{729} + \frac{32}{27} \zeta_3 - \frac{176}{27} \zeta_2 \bigg) + C_A^2 n_f   \bigg(  - \frac{80860}{729} + \frac{704}{9} \zeta_3 + \frac{11960}{81} \zeta_2 - \frac{24}{5} 
         \zeta_2^2 \bigg) + C_A^3   \bigg( \frac{423704}{729} \nonumber\\& + 192 \zeta_5 - \frac{18188}{27} \zeta_3 - \frac{55448}{81} \zeta_2 + 
         \frac{176}{3} \zeta_2 \zeta_3 + \frac{1384}{15} \zeta_2^2 \bigg) + C_F C_A n_f   \bigg(  - \frac{2158}{27} + \frac{472}{9} \zeta_3 + \frac{16}{3} \zeta_2 + \frac{32}{5} \zeta_2^2 \bigg)\,,
\nonumber\\
\varphi^{g,(1)}_{d,3} &=
       C_A n_f^2   \bigg( \frac{56}{27} \bigg) + C_A^2 n_f   \bigg) \frac{1528}{81} - 8 \zeta_3 - \frac{152}{9} \zeta_2 \bigg) + C_A^3   \bigg( - \frac{18988}{81} + \frac{448}{3} \zeta_3 + \frac{752}{9} \zeta_2 \bigg) + C_F C_A n_f   \bigg( 4 - \frac{8}{3} \zeta_2 \bigg).
\nonumber\\
\varphi^{g,(2)}_{d,3} &=
       C_A n_f^2   \bigg( \frac{8}{27} \bigg) + C_A^2 n_f   \bigg( \frac{164}{27} + \frac{2}{3} \zeta_2 \bigg) + C_A^3   \bigg( - \frac{1432}{27} + \frac{40}{3} \zeta_2 \bigg)\,,
\nonumber\\
\varphi^{g,(3)}_{d,3} &= C_A^2 n_f   \bigg( \frac{32}{27} \bigg) + C_A^3   \bigg( - \frac{176}{27} \bigg)
\end{align}
\end{widetext}
The color factors $C_A = N_c$ and $C_F = (N_c^2-1)/2 N_c$ for $\rm{SU}(N_c)$ gauge group. Here, $n_f$ is the number of active flavours and $\zeta_i$ are the Riemann zeta functions. The next task is to systematically resum these SV and NSV logarithms illustrated above, in the threshold region $z_l \rightarrow 1$ where they become numerically large. 

\subsection{Resummation}
We derived the analytical expression of the resummed partonic Coefficient function in article \cite{Ajjath:2020lwb} in double Mellin space where $z_l \rightarrow 1$ translates to large $N_l$ limit. Using the all order intergal representation of $\Psi^g_d$ in Eq.\eqref{eq:psiint} and the RG equation of $a_s$, the Mellin moment of $\Delta^g$ is expressed as,
\begin{align}
\label{eq:delta}
\Delta_{d, N_1, N_2}^g = \tilde g_{d,0}^g \exp(\Psi_{d,N_1, N_2}^g)\,,
\end{align}
where $\Psi_{d,\vec N}^g$ is the double Mellin moment of the function $\Psi^g_d$ and $ \tilde g_{d,0}^g = \sum\limits_{i=0}^{\infty} a_s^i~ \tilde g_{d,0,i}^g $ are the $N$-independent constants. The section 2.2 of \cite{Laenen:2008ux} and appendix A.5 of \cite{Ajjath:2020ulr} contain results required for the computation of Mellin moments of distributions as well as regular terms in the large $N_j$ limit for inclusive cross-section which we extended for rapidity distribution case. Using these results, we computed the resummed result for $\Psi_{d,\vec N}^g$ and it takes the following form,
\begin{align}
\label{eq:PsiN}
\Psi_{d, N_1, N_2}^g = &~~
  g_{d,1}^g(\omega)  \ln N_1
\nonumber\\&
+ \sum\limits_{i=0}^\infty a_s^i \bigg( \frac{1}{2}  g_{d,i+2}^g(\omega) + \frac{1}{N_1} \overline{g}_{d,i+1}^g(\omega) \bigg)
\nonumber\\&
 +\frac{1}{N_1}
\sum\limits_{i=0}^{\infty} a_s^i h^g_{d,i}(\omega,N_1) + (N_1 \leftrightarrow N_2) \,,
\end{align}
where 
\begin{align}
\label{hg}
        h^g_{d,0}(\omega,N_l) &= h^g_{d,00}(\omega) + h^g_{d,01}(\omega) \ln N_l,
        \nonumber\\
         h^g_{d,i}(\omega,N_l) &= \sum_{k=0}^{i} h^g_{d,ik}(\omega)~ \ln^k N_l ,
\end{align}
In the above equation, $\omega = a_s \beta_0 \ln N_1 N_2$ and terms of order ${\cal D}_0 (1/N^i),i>1$ have been dropped. Working in Mellin space has facilitated the entire exponent in Eq.\eqref{eq:delta} to be written in a compact form through $\omega$ dependent functions $g_{d,i}^g$, $\overline{g}_{d,i}^g$ and $h^g_{d,i}$, containing both SV and NSV logarithmic contributions to all orders. Also, the use of resummed $a_s$ allowed us to organise the series in such a way that $\omega$ is treated as order one at every order in $a_s$. The integral representation in $z$ space(Eq.\eqref{eq:psiint}) and the resummed result in Mellin space contain exactly same information regarding SV and NSV contributions, with the only difference that there is no compact looking structure in the former case. The SV resummation coefficients, $\tilde{ g}_{d,0}^g$ and $g_{d,i}^g$ have been discussed in great detail in references \cite{Banerjee:2017cfc,Banerjee:2017ewt,Ahmed:2020caw}. Here, we focus on the NSV resummation coefficients namely $\overline{g}_{d,i}^g$ and $h^g_{d,i}$. The coefficient $\overline{ g}_{d,1}^g$ is found to be identically zero and the remaining coefficients $\overline{g}_{d,i+2}^g$ are functions of universal cusp anomalous dimensions $A^g$. The coefficients $h^g_{d,i}$ depend on the NSV coefficients $\varphi_{d,c}^f$ as well as on $C^g$ , $D^g$ from ${\cal P}^g $. It contains a double series expansion in $a_s(\mu_R^2)$ and $\ln N_l$ and the explicit $\ln N_l$ comes from the explicit $\ln(1-z_l)$ terms in the expansion of $\varphi_{d,g}^f$. The coefficient $h^g_{d,01}$ being proportional to $C^g_1$, is identically zero. Hence, at order $a_s^0$, there is no $(1/N_l)\ln(N_l)$ term. The results for the coefficients $\overline{g}_{d,i}^g$ and $h^g_{d,i}$ are provided in the appendix.

The entire all order information is embedded systematically in the resummation coefficients $\tilde g^g_{d_0,i},g^g_{d,i}(\omega),\overline {g}^g_{d,i}(\omega)$ and $h^g_{d,i}(\omega)$, which can be used to predict SV and NSV terms to all orders. We present Table [\ref{tab:resSV}] and [\ref{tab:resNSV}] below to demonstrates this predictive feature for the SV and NSV terms in  $\Delta_{d,N_1,N_2}^g$ at a given logarithmic accuracy. In Table [\ref{tab:resSV}], we list the predictions for the SV logarithms. The resummed exponents $\{\tilde g^g_{d.0,0}, g^g_{d,1}\}$ given in the first row can predict the leading SV term $a_s^i \ln^l N_1 \ln^k N_2$  with ${l+k=2i}~(l,k \geq 0)$ for all $i>1$ which form the tower of SV-LL resummation. When we include the functions $\{\tilde g^g_{d,0,1},g^g_{d,2}\}$ given in the second row in the exponent $\Psi_{d, N_1, N_2}^g$ along with the first set, we get two towers of next-to-leading SV terms $a_s^i \ln^l N_1 \ln^k N_2$ with $l+k= 2i-1, 2i-2$ for all $i>2$. By using the first and second row resummed exponents, we get the logarithms which constitute SV-NLL resummation. In general, the resummed result with exponents $\{\tilde g^g_{d,0,n},g^g_{d,n+1}\}$ along with the previous sets can predict the term $a_s^i \ln^l N_1 \ln^k N_2$ with $l+k= 2n+1, 2n$ for all $i>n+1$ where $n=0,1,2 \cdots$ and constitutes to the SV-${\rm{NLL}}$ resummation.

The Table [\ref{tab:resNSV}] gives the predictions for NSV logarithms present in $\Delta^{g}_{d,N_1,N_2}$ by including the resummed exponents $\{ \overline g^g_{d,i}$,$h^g_{d,i}\}$ together with the SV resummed exponents.  For instance, using the first set of resummed exponents $\{ \tilde g^g_{d,0,0}, g^g_{d,1}, \overline g^g_{d,1}$, $h^g_{d,0}\}$, we can predict the two leading towers of NSV logarithms $\{a_s^i{\ln^l N_1 \over N_1} \ln^k N_2, a_s^i{\ln^l N_2 \over N_2} \ln^k N_1\}$ with $l+k=2i-1$ for all $i>1$. These two leading towers constitute to the NSV-$\overline{\rm LL}$ resummation. Using the second set of resummed exponents $\{\tilde g^g_{d,0,1},g^g_{d,2},\overline g^g_{d,2}$, $h^g_{d,1}\}$ in addition to the first set give the two towers of next-to-leading NSV terms $\{a_s^i{\ln^l N_1 \over N_1} \ln^k N_2, a_s^i{\ln^l N_2 \over N_2} \ln^k N_1\}$ with $l+k=2i-2$ for all $i>2$ and these towers contribute to the NSV-$\overline{\rm{NLL}}$ resummation. 

\begin{table*}
\centering
\begin{small}
{\renewcommand{\arraystretch}{2.5}
\begin{tabular}{|P{1.6cm}||P{.05cm}P{2.4cm}P{2.9cm}P{2.9cm}P{0.8cm}P{3.5cm}|| P{1.7cm}|}
 \hline
 \hline
 \multicolumn{1}{|c||}{GIVEN} & & \multicolumn{4}{c}{PREDICTIONS - SV Logarithms} & & {Logarithmic Accuracy} \\
 \cline{1-1}\cline{2-7}\cline{1-1}
  Resummed exponents & &$\Delta^{g,(2)}_{d,N_1,N_2}$ & $\Delta^{g,(3)}_{d,N_1,N_2}$&$\Delta^{g,(4)}_{d,N_1,N_2}$& $\cdots$ &$\Delta^{g,(n)}_{d,N_1,N_2}$ &\\
 \hline
 	$\tilde  g^g_{d,0,0},g^g_{d,1}$ &  &  $\Big \{L_1^i L_2^j\Big \}{\Big|}_{i+j=4 }$ & $\Big \{L_1^i L_2^j \Big \}{\Big|}_{i+j=6}$ & $\Big \{L_1^i L_2^j \Big \}{\Big |}_{i+j=8}$ & $\cdots$&$\Big \{L_1^i L_2^j \Big \}{\Big |}_{i+j=2n}$&  ${\rm LL}$ \\
 $\tilde  g^g_{d,0,1},g^g_{d,2}$ &  &  &$\Big \{L_1^i L_2^j \Big \}{\Big |}_{ i+j=5,4}$ &$\Big \{L_1^i L_2^j \Big \}{\Big |}_{ i+j=7,6}$& $\cdots$ &$\Big \{L_1^i L_2^j \Big \}{\Big |}_{ i+j=2n-1,2n-2}$&  ${\rm NLL}$ \\ 
$\tilde  g^g_{d,0,2},g^g_{d,3}$ \newline  &  & & & $\Big \{L_1^i L_2^j\Big \}{\Big |}_{ i+j=5,4}$ & $\cdots$ & $\Big \{L_1^i L_2^j\Big \}{\Big |}_{ i+j=2n-3,2n-4}$  &  ${\rm NNLL}$ \\
 \hline 
 \hline
\end{tabular}}
\caption{\label{tab:resSV} The set of resummed exponents \Big\{$\tilde  g^g_{d,0,n},g^g_{d,n}(\omega)$\Big\} which is required to predict the tower of SV logarithms in $\Delta_{d,N_1,N_2}^{g,(n)}$ at a given logarithmic accuracy in the Mellin $N$-space. Here,  $i,j \geq 0$ and $L^i_l = \ln^i N_l$ with $l=1,2$.}
\end{small}
\end{table*}
\begin{table*}
\centering
\begin{small}
{\renewcommand{\arraystretch}{2.5}
\begin{tabular}{|P{1.6cm}||P{.02cm}P{2.9cm}P{2.9cm}P{2.9cm}P{0.7cm}P{3.5cm}|| P{1.7cm}|}
 \hline
 \hline
 \multicolumn{1}{|c||}{GIVEN} & & \multicolumn{4}{c}{PREDICTIONS - NSV Logarithms} & & {Logarithmic Accuracy} \\
 \cline{1-1}\cline{2-7}\cline{1-1}
  Resummed exponents & &$\Delta^{g,(2)}_{d,N_1,N_2}$ & $\Delta^{g,(3)}_{d,N_1,N_2}$&$\Delta^{g,(4)}_{d,N_1,N_2}$& $\cdots$ &$\Delta^{g,(n)}_{d,N_1,N_2}$ &\\
 \hline
 		\multirow{2}{4em}{$\tilde g^g_{d,0,0},g^g_{d,1}, \newline  \overline g^g_{d,1},h^g_{d,0}$} \newline  &  &  $\Big \{L^{i,j}_{N_{1},2}, L^{i,j}_{N_{2},1} \Big \}{\Big|}_ {i+j=3 }$ & $\Big \{L^{i,j}_{N_{1},2}, L^{i,j}_{N_{2},1} \Big \}{\Big|}_ {i+j=5 }$  &$\Big \{L^{i,j}_{N_{1},2}, L^{i,j}_{N_{2},1} \Big \}{\Big|}_ {i+j=7 }$ &$\cdots$ &$\Big \{L^{i,j}_{N_{1},2}, L^{i,j}_{N_{2},1} \Big \}{\Big|}_ {i+j=2n-1}$&  $\overline{{\rm LL}}$\\
 \multirow{2}{4em}{$\tilde g^g_{d,0,1},g^g_{d,2},\newline \overline g^g_{d,2}, h^g_{d,1}$} \newline   &  &  &$\Big\{L^{i,j}_{N_{1},2}, L^{i,j}_{N_{2},1}\Big\}{\Big|}_ {i+j=4}$ &$\Big\{L^{i,j}_{N_{1},2}, L^{i,j}_{N_{2},1}\Big\}{\Big|}_ {i+j=6}$& $\cdots$ &$\Big \{L^{i,j}_{N_{1},2}, L^{i,j}_{N_{2},1} \Big \}{\Big|}_ {i+j=2n-2}$&  $\overline{{\rm NLL}}$ \\ 

\multirow{2}{4em}{$\tilde g^g_{d,0,2},g^g_{d,3},\newline \overline g^g_{d,3}, h^g_{d,2}$} \newline \newline   &  & & & $\Big \{L^{i,j}_{N_{1},2}, L^{i,j}_{N_{2},1}\Big \}{\Big|}_ {i+j=5}$ & $\cdots$ & $\Big \{L^{i,j}_{N_{1},2}, L^{i,j}_{N_{2},1} \Big \}{\Big|}_ {i+j=2n-3}$  &  $\overline{{\rm NNLL}}$ \\
 \hline 
 \hline
\end{tabular}}
\caption{\label{tab:resNSV} The set of resummed exponents \Big\{$\tilde g^g_{d,0,n},g^g_{d,n}(\omega),\overline{g}^g_{d,n}(\omega), h^g_{d,n}(\omega)$\Big\} which is required to predict the tower of NSV logarithms in $\Delta_{d,N_1,N_2}^{g,(n)}$ at a given logarithmic accuracy in the Mellin $N$-space. Here, $i,j \geq 0$, $L^{i,j}_{N_{1},2} = {\ln^i N_1 \over N_1} \ln^j N_2$ and $L^{i,j}_{N_{2},1} = {\ln^i N_2 \over N_2} \ln^j N_1$ .}
\end{small}
\end{table*}


\begin{widetext}
\subsection{All order Prediction}

In the previous section, we showed that using a particular set of resumed exponents at each logarithmic accuracy, we can predict certain SV and NSV logarithms in $\Delta^{g,(n)}_{d,N_{1},N_{2}}$ at each order in the perturbation theory in the Mellin N-space. This is depicted in Table [\ref{tab:resSV}] and [\ref{tab:resNSV}] given above. Here, we present those resulting predictions for the NSV logarithms till fourth order in perturbation theory using the set of resumed exponents belonging to $\rm \overline{LL}$, $\rm \overline{NLL}$ and $\rm \overline{NNLL}$ resumed accuracy respectively as shown in Table [\ref{tab:resNSV}]. Our results do not have renormalisation and factorisation scales explicitly as we have set $\mu_R = \mu_F = m_H$. The SV+NSV resumed result at the leading logarithmic($\rm \overline{LL}$) accuracy is given by the expression,
\begin{align}
\label{LL1}
    \Delta_{d,N_1,N_2}^{g,\overline{\rm{LL}}} &= \tilde g^g_{d,0,0} \exp\bigg[\ln N_1 ~ g^g_{d,1}(\omega)+ \frac{1}{N_1}\bigg(\overline g^g_{d,1}(\omega) +  h^g_{d,0}(\omega,N_1)\bigg)\bigg] + (N_1 \leftrightarrow N_2)\,.
\end{align}

The resumed exponents present in the above equation depend on one loop anomalous dimensions and process dependent finite coefficients obtained from fixed order NLO results. Thus, the above equation provides leading SV and NSV logarithms at every order in perturbation theory using only one loop information. Below, we give the leading NSV logarithms at $a_s^2 (\rm NNLO)$ , $a_s^3 (\rm N^3LO)$  and $a_s^4 (\rm N^4LO)$ level resulting from the $\rm \overline{LL}$ resummation expression. At $a_s^2$ (NNLO), we get
\begin{align}
   \Delta_{d,N_1,N_2}^{g,(2)}|_{\rm{NSV}-\overline{\rm{LL}}}   &=   
       L^3_{N_{1}}  \bigg\{
           4  C_A^2
          \bigg\}
       +   L^{2,1}_{N_{1},2}  \bigg\{
           12  C_A^2
          \bigg\}
       + L^{1,2}_{N_{1},2}  \bigg\{
           12  C_A^2
          \bigg\}
       + {L^3_2\over N_1 }   \bigg\{
           4  C_A^2
          \bigg\}+ (N_1 \leftrightarrow N_2)  \,.
    \end{align}
At $a_s^3$ (N$^3$LO), we obtain
\begin{align}
   \Delta_{d,N_1,N_2}^{g,(3)}|_{\rm{NSV}-\overline{\rm{LL}}}   &=   
          L^5_{N_{1}}  \bigg\{
           4 C_A^3
          \bigg\}
       + L^{4,1}_{N_{1},2}     \bigg\{
           20 C_A^3
          \bigg\}
           +  L^{3,2}_{N_{1},2}  \bigg\{
           40 C_A^3
          \bigg\}
       +  L^{2,3}_{N_{1},2}  \bigg\{
           40 C_A^3
          \bigg\}
       +L^{1,4}_{N_{1},2} \bigg\{
           20 C_A^3
          \bigg\}
       + {L^5_2 \over N_1}   \bigg\{
           4 C_A^3 
          \bigg\} \\ \nonumber &
        +  (N_1 \leftrightarrow N_2)  \,.
    \end{align}
Expanding Eq.(\ref{LL1}) upto $a_s^4$ (N$^4$LO) gives, 
\begin{align}
   \Delta_{d,N_1,N_2}^{g,(4)}|_{\rm{NSV}-\overline{\rm{LL}}}   &=   
     L^7_{N_{1}}  \bigg\{
           {8 \over 3} C_A^4
          \bigg\}

       + L^{6,1}_{N_{1},2}     \bigg\{
           {56 \over 3} C_A^4
          \bigg\}
       +  L^{5,2}_{N_{1},2}   \bigg\{
           56 C_A^4
          \bigg\}
       +L^{4,3}_{N_{1},2}   \bigg\{
           {280 \over 3} C_A^4
          \bigg\}
        +L^{3,4}_{N_{1},2}   \bigg\{
           {280 \over 3} C_A^4
          \bigg\}  \\ \nonumber &
        +L^{2,5}_{N_{1},2}   \bigg\{
           56 C_A^4
          \bigg\}
           +L^{1,6}_{N_{1}}   \bigg\{
           {56 \over 3} C_A^4
          \bigg\}
        + {L^7_2 \over N_1}   \bigg\{
           {8 \over 3} C_A^4 
          \bigg\}+ (N_1 \leftrightarrow N_2)  \,, 
    \end{align}
where $L^{i,j}_{N_{1},2} = {\ln^i N_1 \over N_1} \ln^j N_2$, $L^k_{N_l} = {\ln^k N_l \over N_l}$ and $L^k_l = \ln N_l$ with $l=1,2$. These predictions for the leading NSV logarithms have been compared with the fixed-order results given in Refs.\cite{Anastasiou:2004xq,Dulat:2018bfe} till third order. \\

We next look at the expression for the SV+NSV resumed result at next-to-leading logarithmic($\rm \overline{NLL}$) accuracy given below, 
\begin{align}
\label{NLL}
    \Delta_{d,N_1,N_2}^{g,\overline{\rm{NLL}}} &= ( \tilde g^g_{d,0,0} + a_s ~ \tilde g^g_{d,0,1} )~\exp\bigg[\ln N_1 ~ g^g_{d,1}(\omega) 
    +  g^g_{d,2}(\omega)   
    + \frac{1}{N_1}\bigg(\overline g^g_{d,1}(\omega) +  a_s ~\overline g^g_{d,2}(\omega)
  +  h^g_{d,0}(\omega,N_1)  \\ \nonumber & + a_s ~  h^g_{d,1}(\omega,N_1) \bigg)\bigg]  + (N_1 \leftrightarrow N_2) \,.
 \end{align}
At $\rm \overline{NLL}$ accuracy, the resumed exponents depend on the anomalous dimensions expanded upto two loops and the finite process dependent SV and NSV coefficients obtained from fixed order results upto NNLO accuracy. The above equation provides next-to-leading SV and NSV logarithms at every order in perturbation theory using information embedded in 2-loop results. The next-to-leading NSV logarithms at $a_s^3$ (N$^3$LO) are given as,
\begin{align}
  \Delta_{d,N_1,N_2}^{g,(3)}|_{\rm{NSV}-\overline{\rm{NLL}}}   &~=     \Delta_{d,N_1,N_2}^{g,(3)}|_{\rm{NSV}-\overline{\rm{LL}}} 
    + L^4_{N_1} \bigg\{ + \Big(\frac{326}{9} + 40 ~ \gamma_E \Big) C_A^3 \bigg\} 
     - \frac{20}{9} C_A^2 n_f \bigg \}
    + L^{3,1}_{N_1,2} \bigg\{ \Big(  \frac{1160}{9} + 160 \gamma_E \Big) C_A^3 \\ \nonumber &
     - {80 \over 9} C_A^2 n_f  \bigg\}
    + L^{2,2}_{N_1,2}  \bigg\{ \Big( {484 \over 3}  + 240 \gamma_E \Big) C_A^3 
     - {40 \over 3} C_A^2 n_f  \bigg\}
    + L^{1,3}_{N_1,2}  \bigg\{ \Big( {728 \over 9}  + 160 \gamma_E \Big) C_A^3 - {80 \over 9} C_A^2 n_f  \bigg\} \\ \nonumber &
     + {L^4_2 \over N_1} \bigg\{  \Big( {110 \over 9}  + 40 \gamma_E \Big) C_A^3 
     - {20 \over 9} C_A^2 n_f  \bigg\}
    + (N_1 \leftrightarrow N_2) \,,
\end{align}
At $a_s^4$ (N$^4$LO), we find
\begin{align}
  \Delta_{d,N_1,N_2}^{g,(4)}|_{\rm{NSV}-\overline{\rm{NLL}}}   &~=     \Delta_{d,N_1,N_2}^{g,(4)}|_{\rm{NSV}-\overline{\rm{LL}}} 
    + L^6_{N_1} \bigg\{ \Big(\frac{370}{9} + {112 \over 3} ~ \gamma_E \Big) C_A^4  
     - \frac{28}{9} C_A^3 n_f 
     \bigg\}
     
    + L^{5,1}_{N_1,2}  \bigg\{  \Big( \frac{692}{3} + 224 \gamma_E \Big) C_A^4 \\ \nonumber &
     - {56 \over 3} C_A^3 n_f  \bigg\}
     
    + L^{4,2}_{N_1,2} \Bigg\{ \Big( \frac{1586}{3}  + 560 \gamma_E \Big) C_A^4 
     - {140 \over 3} C_A^3 n_f  \bigg\}
     
    + L^{3,3}_{N_1,2}  \bigg\{ \Big( \frac{5672}{9} + {2240 \over 3} \gamma_E \Big) C_A^4  \\ \nonumber &
     - {560 \over 9} C_A^3 n_f  \bigg\}
     
    + L^{2,4}_{N_1,2}  \bigg\{ \Big(  \frac{1226}{3} + 560 \gamma_E \Big) C_A^4 
     - {140 \over 3} C_A^3 n_f \bigg\}

    + L^{1,5}_{N_1,2}  \bigg\{ \Big(  \frac{404}{3} + 224 \gamma_E \Big) C_A^4 \\ \nonumber &
     - {56 \over 3} C_A^3 n_f  \bigg\}
     
    + {L^6_2 \over N_1} \bigg\{ {154 \over 9}  C_A^4 
     - {28 \over 9} C_A^3 n_f + {112 \over 3} \gamma_E  C_A^4 \bigg\}
    + (N_1 \leftrightarrow N_2) \,,
    \end{align}
where $\gamma_E$ is the Euler-Mascheroni constant. These predictions for the next-to-leading NSV logarithms are in agreement with results given in Refs.\cite{Anastasiou:2004xq,Dulat:2018bfe}. \\

At last, we provide the predictions resulting from the $\rm \overline{NNLL}$ resummation. The expression used to obtain the results at this logarithmic accuracy is given as,
\begin{align}
\label{NNLL}
    \Delta_{d,N_1,N_2}^{g,\overline{\rm{NNLL}}} &= ( \tilde g^g_{d,0,0} + a_s ~ \tilde g^g_{d,0,1} +  a_s^2 ~ \tilde g^g_{d,0,2})
    \exp\bigg[\ln N_1 ~ g^g_{d,1}(\omega)
    +  g^g_{d,2}(\omega) + a_s ~ g^g_{d,3}(\omega) 
    + \frac{1}{N_1}\bigg(\overline g^g_{d,1}(\omega) +  a_s ~\overline g^g_{d,2}(\omega) 
     \\ \nonumber & + a_s^2 ~\overline g^g_{d,3}(\omega) 
    +  h^g_{d,0}(\omega,N_1) + a_s ~  h^g_{d,1}(\omega,N_1) 
    +  a_s^2 ~  h^g_{d,2}(\omega,N_1) \bigg)\bigg] + (N_1 \leftrightarrow N_2)  \,.
\end{align}
      
We note that the $\rm \overline{NNLL}$ resummation encapsulates three loop information coming from anomalous dimensions expanded till third order in strong coupling constant and also from the third order SV and NSV finite coefficients obtained from $\rm N^3LO$ results. We provide the prediction for the next-to-next-to leading NSV logarithm at $a_s^4$ as given below,  
\begin{align}

\Delta_{d,N_1,N_2}^{g,(4)}|_{\rm{NSV}-{\overline{\rm{NNLL}}}}   &=    

\Delta_{d,N_1,N_2}^{g,(4)}|_{\rm{NSV}-{\overline{\rm{NLL}}}} 

+ L^5_{N_1} \bigg\{ 
  {32 \over 27} C_A^2 n_f^2  
- \Big ({1172 \over 27} + {112 \over 3}\gamma_E \Big ) C_A^3 n_f 
+ \Big (\frac{2024}{9} + \frac{1432}{3} \gamma_E + 224 \gamma_E^2 \\ \nonumber &
+ 40 \zeta_2 \Big ) C_A^4 \bigg\} 

+ L^{4,1}_{N_1,2}  \bigg\{
  {160 \over 27} C_A^2 n_f^2 
- \Big({5608 \over 27} + {560 \over 3}\gamma_E \Big) C_A^3 n_f 
+ \Big(\frac{9604}{9} + \frac{6632}{3} \gamma_E + 1120 \gamma_E^2  \\ \nonumber &
+ 200 \zeta_2 \Big) C_A^4 \bigg\}
   
+  L^{3,2}_{N_1,2}  \bigg\{
  {320 \over 27} C_A^2 n_f^2 
- \Big({10568 \over 27 } + {1120 \over 3}\gamma_E \Big) C_A^3 n_f 
+ \Big(\frac{17912}{9} + \frac{12016}{3} \gamma_E + 2240 \gamma_E^2  \\ \nonumber &
+ 400 \zeta_2 \Big) C_A^4  \bigg\}
  
+ L^{2,3}_{N_1,2} \bigg\{
  {320 \over 27} C_A^2 n_f^2   
- \Big(\frac{9712}{27} + {1120 \over 3}\gamma_E \Big) C_A^3 n_f 
+ \Big(\frac{48704}{27} + \frac{10576}{3} \gamma_E + 2240 \gamma_E^2 \\ \nonumber &
+ 400 \zeta_2 \Big) C_A^4  \bigg\}  
  
+ L^{1,4}_{N_1,2}  \bigg\{
  {160 \over 27} C_A^2 n_f^2 
- \Big(\frac{4292}{27} + {560 \over 3}\gamma_E \Big) C_A^3 n_f 
+ \Big(\frac{21088}{27} + \frac{4472}{3} \gamma_E + 1120 \gamma_E^2 \\ \nonumber &
+ 200 \zeta_2 \Big) C_A^4  \bigg\}

+ \frac{L^5_2}{N_1} \bigg\{
  {32 \over 27} C_A^2 n_f^2 
- \Big ({712 \over 27} + {112 \over 3}\gamma_E \Big ) C_A^3 n_f 
+ \Big (\frac{3380}{27} + \frac{712}{3} \gamma_E + 224 \gamma_E^2 + 40 \zeta_2 \Big ) C_A^4 \bigg\}\,.

\end{align}

\end{widetext}

The predictions given in this section, for the NSV logarithms in the rapidity distribution $\Delta^{g,(i)}_{d,N_1,N_2}$ at different logarithmic accuracy can reproduce the corresponding predictions for the inclusive cross-section $\Delta^{g,(i)}_{N}$ in the limit $N_1 = N_2 = N$ and has been verified till fourth order using Ref.\cite{Ajjath:2021bbm}.

\section{Phenomenology}
\begin{table*}[ht]
 \renewcommand{\arraystretch}{1.9}
\begin{tabular}{ |P{1.0cm}||P{1.3cm}|P{1.1cm}|P{1.5cm} |P{1.2cm}|P{1.9cm}|}
 \hline
  y &$\rm{K_{LO+\rm{\overline{LL}}}}$&$\rm{K_{NLO}}$&$\rm{K_{NLO+\rm{\overline{NLL}}}}$&$\rm{K_{NNLO}}$&$\rm{K_{NNLO+\rm{\overline{NNLL}}}}$\\
 \hline
\hline
 0-0.4    & 1.775 & 1.782 & 1.854 & 2.092 & 2.030 \\
 0.4-0.8  & 1.776 & 1.755 & 1.828 & 2.079 & 2.019  \\
 0.8-1.2  & 1.796 & 1.725 & 1.804 & 2.031 & 1.973 \\
 1.2-1.6  & 1.812 & 1.679 & 1.763 & 1.959 & 1.904 \\
 1.6-2.0  & 1.853 & 1.616 & 1.711 & 1.897 & 1.849 \\
 2.0-2.4  & 1.891 & 1.535 & 1.640 & 1.794 & 1.752 \\
\hline
\end{tabular}
\caption{K-factor values of fixed order and resummed results at the central scale $\mu_R=\mu_F= m_H/2$.} \label{tab:KmZ}
\end{table*}
This section presents a detailed study on the numerical impact of resummed soft virtual plus next-to-soft virtual (SV+NSV) corrections upto $\rm \overline{NNLL}$ accuracy for the rapidity distribution of the Higgs boson production in gluon fusion at the LHC. To distinguish between the SV and SV+NSV resummed results, the NSV included resummed results have been denoted by $\rm \overline{N^n LL}$ for the $n^{th}$ level logarithmic accuracy. We do the analysis for centre of mass energy $\sqrt{S} = 13$ TeV with the Higgs mass $m_H = 125$ GeV, top quark mass $m_t = 173.2$ GeV and Fermi constant $G_F = 4541.63$ pb. The fixed order rapidity distributions have been obtained using a publicly available code FEHiP. An in-house FORTRAN code has been used to perform double Mellin inversion for the resummed contributions. We have used Minimal prescription \cite{Catani:1996yz} to deal with the Landau pole in the Mellin inversion routines. The PDFs used are taken from the \textbf{LHAPDF} \cite{Buckley:2014ana} routine using the \textbf{MMHT2014(68cl)} \cite{Harland-Lang:2014zoa} parton distribution set. The strong coupling constant $a_s$ is provided using the \textbf{LHAPDF} interface with $n_f = 5$ active massless quark flavours throughout. The Mellin space PDFs ($f_{i,N}$) can be obtained by using QCD-PEGASUS \cite{Vogt:2004ns}. However, we follow the technique given in \cite{Catani:1989ne,Catani:2003zt} to directly deal with PDFs in the $z$ space. The resummed results are matched to the fixed order result in order to avoid any double counting of threshold logarithms and is given as,
\begin{widetext}
\begin{align}\label{match}
{d \sigma^{g,\rm {N^nLO+\overline {\rm N^nLL}}} \over  dy } = &  
{d \sigma^{g, \rm {N^nLO}} \over  dy } +
\, {\sigma^{g}_B }  \int_{c_{1} - i\infty}^{c_1 + i\infty} \frac{d N_{1}}{2\pi i}
 \int_{c_{2} - i\infty}^{c_2 + i\infty} \frac{d N_{2}}{2\pi i} 
\left({\tau}\right)^{-N_{1}-N_{2}} 
\delta_{a b}f_{a,N_1}(\mu_F^2) f_{b,N_2}(\mu_F^2) \\ \nonumber &
\times \bigg( \Delta_{d,N_1,N_2}^g \bigg|_{\overline {\rm {N^nLL}}} - {\Delta^g_{d,N_1,N_2}}\bigg|_{tr ~\rm {N^nLO}} \bigg) \,,
\end{align}
\end{widetext}
Here, $\sigma^g_B$ is the Born prefactor and the contour $c_i$ in the Mellin inversion can be chosen according to {\it Minimal prescription} \cite{Catani:1996yz} procedure. The first term in the Eq. \ref{match} encapsulates the fixed order contributions upto $\rm N^n LO$ accuracy and the second term corresponds to the resummed result at $\rm \overline{N^n LL}$ accuracy obtained by taking the difference between the resummed result and the same truncated upto order $a_s^n$. Therefore, the second term contains the SV+NSV resummed contributions to all orders in perturbation theory starting from $a^{n+1}_s$ onwards.

We have calculated the percentage contribution of SV distributions and NSV logarithms to the Born cross section at various orders in Table 2 of Ref. \cite{Ajjath:2021lvg} at the central scale $\mu_R = \mu_F = m_H$ for inclusive cross section of
Higgs Boson production in gluon fusion. Similar percentages have been computed for Drell-Yan process as well in the Ref.\cite{Ajjath:2021bbm} at the central scale $\mu_R = \mu_F = 200$ GeV. Our analysis showed that for both the cases, the phenomenological relevance of NSV contribtuions increase with each order in the perturbation theory. In the case of Higgs boson production, $gg$-channel being the dominant contributor, gives us an additional motivation to study NSV logarithms as our formalism resums these collinear logarithms coming from diagonal channel only. Recently, we carried out the phenomenological study for the rapidity distribution of Drell-yan process \cite{Ajjath:2021pre} and demonstrated that the NSV contribution plays an important role for rapidity distribution as well. Similarly, we expect the same trends to follow for the rapidity distribtuion of Higgs Boson production through gluon fusion. Now, we ask the following questions to understand the importance of NSV terms and also to shed some light on the role of beyond NSV terms in the rapidity distribution of Higgs Boson production.
\begin{itemize}
\item
How is the behaviour of Fixed order rapidity distribution altered with the inclusion of SV+NSV resummed terms? 
\item
What is the change in the sensitivity of the result on unphysical scales $\mu_R$ and $\mu_F$ when NSV logarithms from dominant $gg$-channel are included ?
\item 
How much is the impact of SV+NSV resummed results on the rapidity distribution in comparison to the well established SV predictions ?
\end{itemize}

We will explore the above questions in the subsequent sections. We have done all the analysis for central scale $\mu_R = \mu_F = m_H/2$ at 13 TeV LHC. The resummation scheme chosen throughout the paper is $\overline{N}$ exponentiation. This choice of central scale and resummation scheme is inspired from the analysis done in Ref.\cite{Ajjath:2021lvg} for the inclusive cross-section of the Higgs boson production through gluon fusion. Also, the fixed order results contain contributions from all the channels whereas the resummed results contain distributions and logarithms coming from the diagonal $gg$ channel only. The NSV logarithms resulting from off-diagonal $qg$ channel are not included in our formalism. Let us start the next section by analysing the effect of SV+NSV resummed result on the fixed order predictions for the rapidity distribution. 

\subsection{Fixed order vs resummed results}
\begin{figure*}
\centering
\includegraphics[scale=0.40]{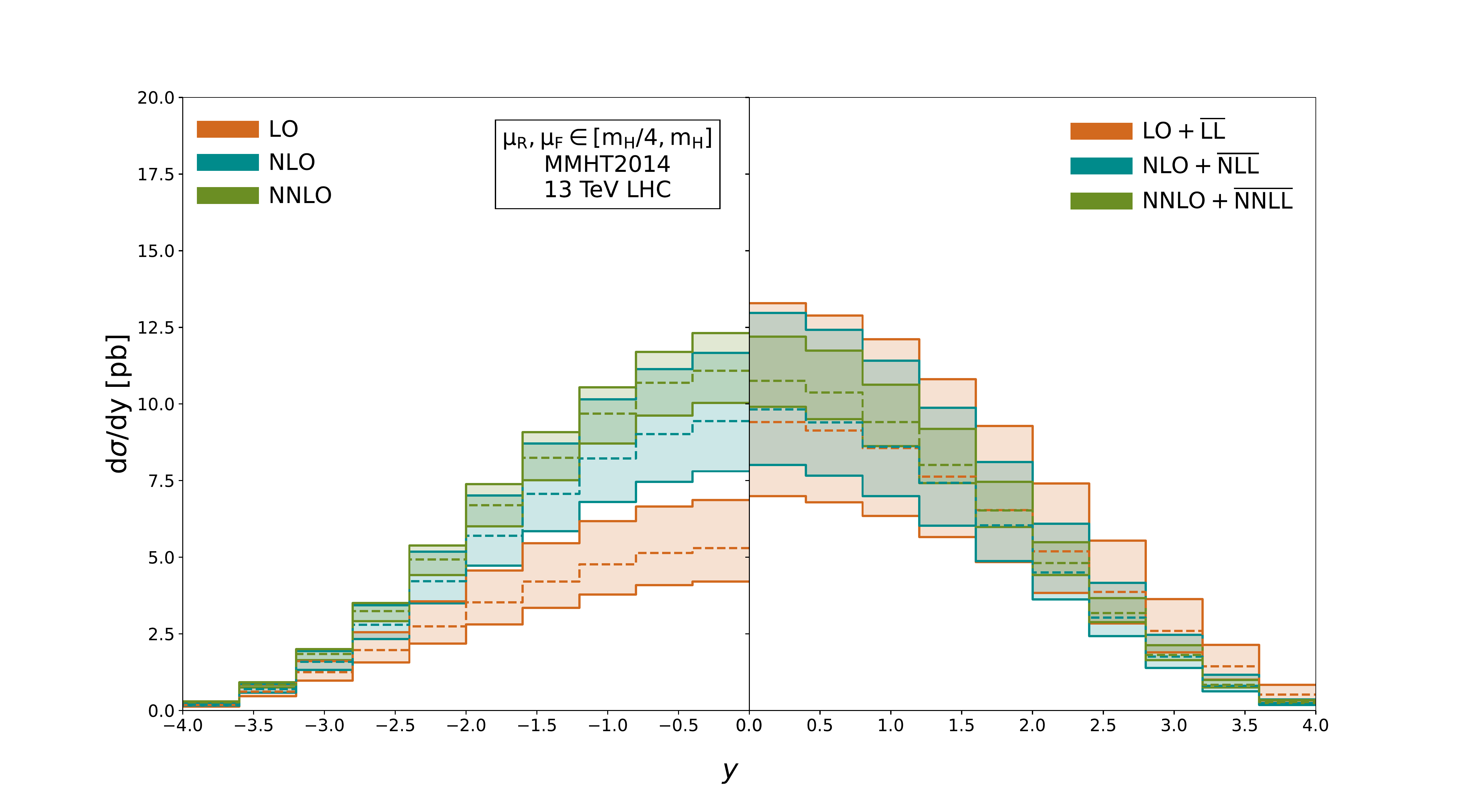}
\caption{Comparison of 7-point scale variation between fixed order and SV+NSV resummed results for 13 TeV LHC. The dashed lines refer to the corresponding central scale
values at each order.}
\label{fig:7ptfixedsv}
\end{figure*}

\begin{table*}
 \renewcommand{\arraystretch}{2.7}
\begin{tabular}{ |P{1.0cm}||P{2.2cm}|P{2.2cm}|P{2.2cm}||P{2.2cm}|P{2.2cm}|P{2.2cm}|}
 \hline
  y &LO&LO+$\rm{\overline{LL}}$&NLO&NLO+$\rm{\overline{NLL}}$&NNLO&NNLO+$\rm{\overline{NNLL}}$\\
 \hline
\hline
 0-0.4 & 5.2987 $^{+ 1.57 }_{- 1.089 }$ 
 & 9.406 $^{+ 3.878 }_{- 2.411 }$ 
 & 9.4432 $^{+ 2.223 }_{- 1.639 }$ 
 & 9.8252 $^{+ 3.144 }_{- 1.816 }$ 
 & 11.0873 $^{+ 1.230 }_{- 1.051 }$ 
 & 10.7559 $^{+ 1.437 }_{- 0.849 }$  \\
 
 0.4-0.8 & 5.1408 $^{+ 1.509 }_{- 1.054 }$ 
 & 9.1303 $^{+ 3.752 }_{- 2.338 }$ 
 & 9.0205 $^{+ 2.121 }_{- 1.563 }$ 
 & 9.3991 $^{+ 3.023 }_{- 1.739 }$ 
 & 10.6907 $^{+  1.011 }_{- 1.068 }$ 
 & 10.3777 $^{+ 1.361 }_{- 0.872 }$ \\
 
 0.8-1.2 & 4.7667 $^{+ 1.407 }_{- 0.988 }$ 
 & 8.561 $^{+ 3.548 }_{- 2.211 }$ 
 & 8.2239 $^{+ 1.925 }_{- 1.422 }$ 
 & 8.601 $^{+ 2.815 }_{- 1.608 }$ 
 & 9.6837 $^{+ 0.855 }_{- 0.969 }$ 
 & 9.407 $^{+ 1.222 }_{- 0.783 }$ \\

 1.2-1.6 & 4.2089 $^{+ 1.248 }_{- 0.864 }$ 
 & 7.626 $^{+ 3.179 }_{- 1.967 }$ 
 & 7.0677 $^{+  1.645 }_{- 1.217 }$ 
 & 7.4229 $^{+ 2.457 }_{- 1.393 }$ 
 & 8.2437 $^{+ 0.836 }_{- 0.736 }$ 
 & 8.0143 $^{+ 1.170 }_{- 0.594 }$  \\

 1.6-2.0 & 3.5278 $^{+ 1.035 }_{- 0.725 }$ 
 & 6.537 $^{+ 2.749 }_{- 1.70  }$ 
 & 5.6996 $^{+ 1.316 }_{- 0.976 }$ 
 & 6.036 $^{+ 2.074 }_{- 1.159 }$ 
 & 6.6937 $^{+ 0.689 }_{- 0.681 }$ 
 & 6.524 $^{+ 0.937 }_{- 0.537 }$ \\
 
 2.0-2.4 & 2.7461 $^{+ 0.817 }_{- 0.564 }$ 
 & 5.1918 $^{+ 2.213 }_{- 1.358 }$ 
 & 4.2150 $^{+ 0.963 }_{- 0.717 }$ 
 & 4.5031 $^{+ 1.592 }_{- 0.878 }$ 
 & 4.9262 $^{+ 0.453 }_{-  0.505 }$ 
 & 4.8123 $^{+ 0.677 }_{- 0.390 }$  \\
\hline
\end{tabular}
\caption{Values of resummed rapidity distribution at various orders in comparison to the fixed order results in pb at the central scale $\mu_R=\mu_F= m_H/2$ for 13 TeV LHC.} \label{tab:FONSV7ptmZ}
\end{table*}

This section presents the detailed study on the numerical relevance of SV+NSV resummed contributions at $\rm \overline{LL}$, $\rm \overline{NLL}$ and $\rm \overline{NNLL}$ matched with the corresponding fixed order results for the rapidity distribution using Eq.\ref{match}. We investigate the enhacement of the SV+NSV resummed matched result with the fixed order counterparts. We also demonstrate the impact of resummation of NSV logarithms from the diagonal channel on the sensitivity of fixed order result w.r.t unphysical scales $\mu_R$ and $\mu_F$.

We begin by studying the quantitative impact of SV+NSV resummed rapidity distribution through the K-factors defined as, 
\begin{equation}\label{eq:Kfac}
 \mathrm{K}  = \dfrac{\dfrac{d\sigma}{ dy}\left(\mu_R=\mu_F=m_H/2\right)}{\dfrac{d\sigma^{\text{LO}}}{ dy}(\mu_R=\mu_F=m_H/2)} 
 \end{equation}
where the renormalisation($\mu_R$) and factorisation($\mu_F$) scales have been set at $m_H/2 = 62.5$ GeV. We provide a Table (\ref{tab:KmZ})
to present K-factor values of fixed order as well as SV+NSV resummed results for benchmark rapidity values. We find that there is an enhancement of 77.5\% and 4.04\% when the resummed SV+NSV logarithms at $\rm \overline{LL}$ and $\rm \overline{NLL}$ are added to LO and NLO respectively at the central rapidity region. However, the rapidity distribution decreases by 2.96\% when we include $\rm \overline{NNLL}$ to NNLO at the central rapidity region. This suggests better perturbative convergence at NNLO+$\rm \overline{NNLL}$ level. We further observe that the percentage enhancement in the rapidity distribution at NNLO+$\rm \overline{NNLL}$ over NLO+$\rm \overline{NLL}$ is less than the enhancement when we go from NLO to NNLO accuracy for a wide range of rapidity values. This indicates that the inclusion of resummed SV+NSV logarithms make the perturbative predictions more reliable.
The above analysis based on the K-factor values demonstrates that the SV+NSV resummed results bring substantial percentage correction to the fixed order results and also improve the perturbative convergence and reliability of the predictions. Now, we proceed to see the dependence of resummed results on the renormalisation and factorisation scales.

\subsection{\textit{7-point scale variation of the resummed result}}
The truncation of the perturbative series to certain order of accuracy plagues the fixed order as well as resummed predictions with the dependence on unphysical scales namely renormalisaion ($\mu_R$) and factorisation ($\mu_R$) scales. Here, we assess the change in uncertainty w.r.t these unphysical scales when the SV+NSV resummed contributions are added to the fixed order results. We use the standard canonical 7-point variation approach where $\{ \mu_R, \mu_F \}$ is varied in the range $\{ m_H/4, m_H \}$,  keeping the ratio $\frac{\mu_R}{\mu_F}$ not larger than 2 and smaller than 1/2. Fig.(\ref{fig:7ptfixedsv}) shows the bin-integrated rapidity distribution of the Higgs boson for fixed order as well as SV+NSV resummed predictions at various orders in perturbation theory. We plot the 7-point scale uncertainties of the fixed order result upto NNLO in the left panel and for the SV+NSV resummed predictions upto NNLO+ $\rm \overline{NNLL}$ in the right panel around the central scale $\mu_R = \mu_F = m_H/2$ for 13 TeV LHC. We observe that there is a significant enhancement of 77.51\% and 4.05\% at LO and NLO accuracy by the addition of $\rm \overline{LL}$ and $\rm \overline{NLL}$ contributions at the central rapidity region. However, the inclusion of $\rm \overline{NNLL}$ result decreases the rapidity distribution at NNLO level by 2.99\% at the central rapidity region. This hints towards better perturbative convergence and improved reliability of the perturbative series as suggested before by K-factor analysis. 


\begin{figure*}
\includegraphics[scale=0.40]{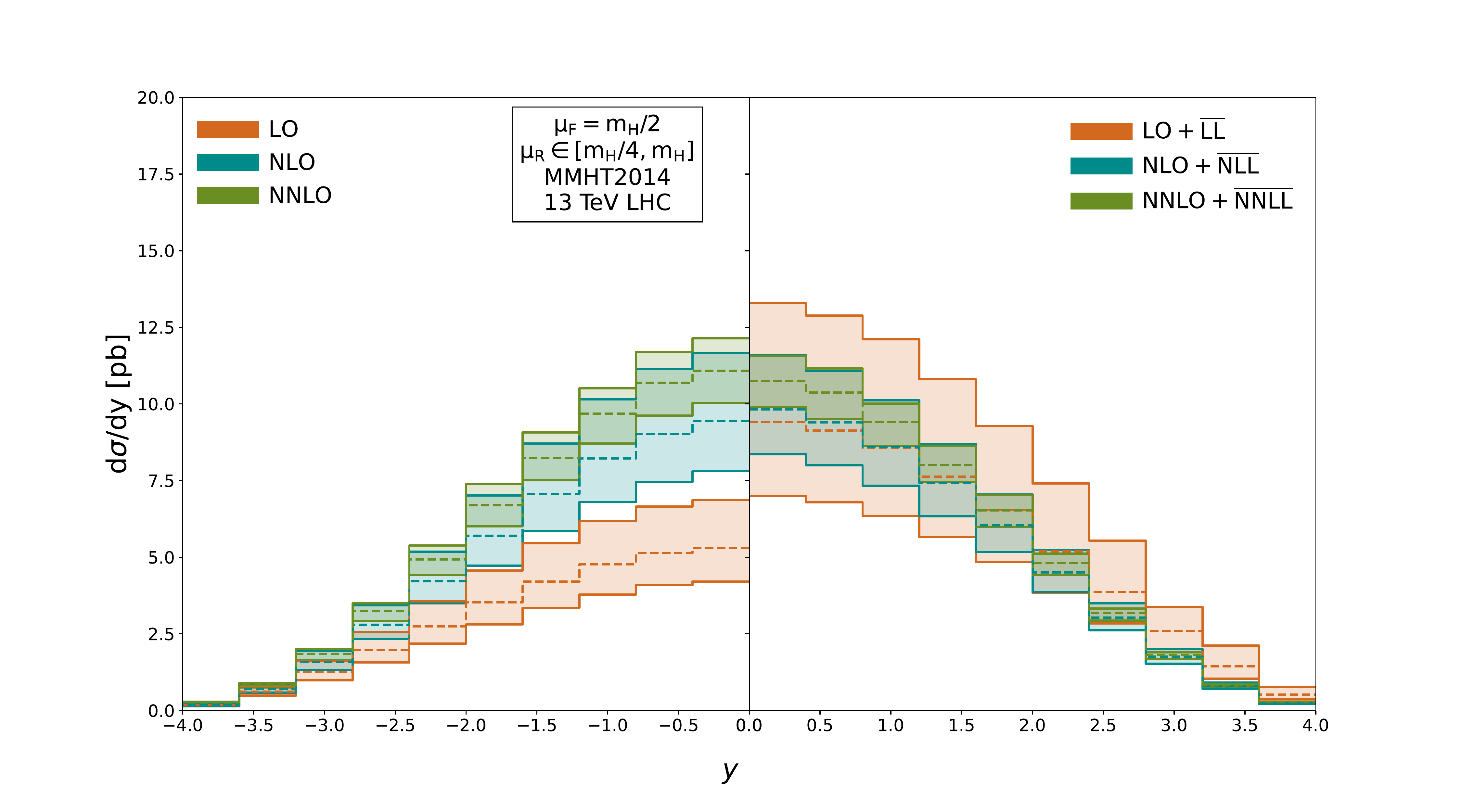}
\caption{Comparison of $\mu_R$ scale variation between fixed order and SV+NSV resummed results with the scale $\mu_F = m_H/2$. The dashed lines refer to the corresponding central scale
values at each order.}
\label{fig:muFfixednsv}
\end{figure*}

\begin{figure*}
\includegraphics[scale=0.40]{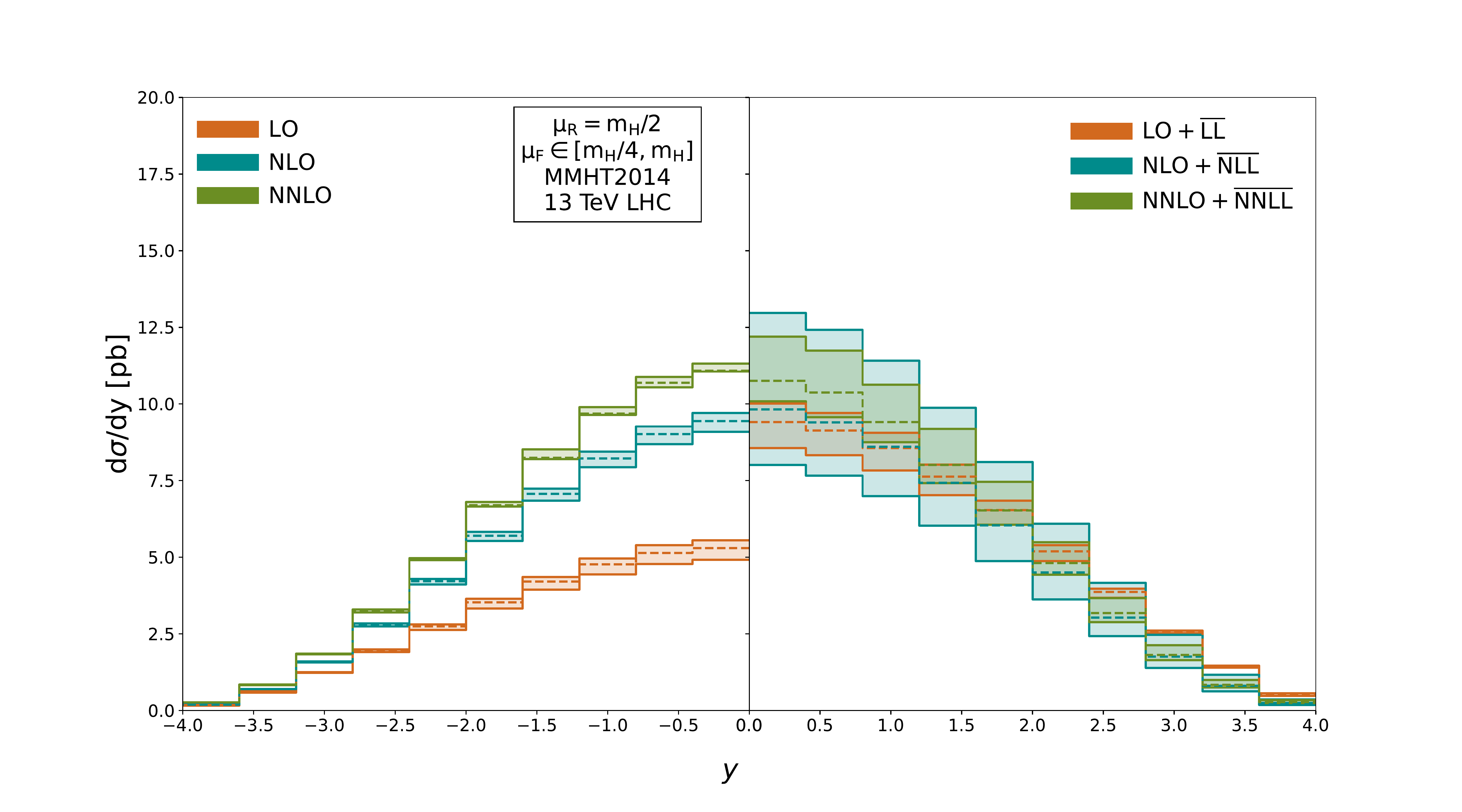}
\caption{Comparison of $\mu_F$ scale variation between SV and SV+NSV resummed results with the scale $\mu_R = m_H/2$. The dashed lines refer to the corresponding central scale
values at each order.}
\label{fig:muRfixednsv}
\end{figure*}
We now compare the 7-point uncertainties of SV+NSV resummed results with the fixed order results for the rapidity distribution at various orders. From Table \ref{tab:FONSV7ptmZ}, we find that the combined uncertainty due to $\mu_R$ and $\mu_F$ varies between (+11.09\%, -9.48\%) at NNLO accuracy which is a substantial reduction as compared to the uncertainty of (+29.63\%, -20.55\%) at LO for the central rapidity region which is as expected. Fig.(\ref{fig:7ptfixedsv}) shows that the uncertainty bands for the resummed results are visibly wider in comparison to the corresponding fixed order predictions till NLO accuracy. The uncertainty varies between (+29.63\%, -20.55\%) for LO whereas it lies between (+41.23\%, -25.63\%) for LO+$\rm \overline{LL}$ for central rapidity region. Similarly, the combined scale uncertainty ranges between (+32\%, -18.48\%) for NLO + $\rm \overline{NLL}$ which is higher than (+23.54\%, -17.36\%) for NLO accuracy around $y=0$. On the contrary, at NNLO accuracy, we notice that the uncertainty bands become comparable for fixed order and SV+NSV resummed predictions. For instance, the 7-point scale uncertainty lies in the range (+11.09\%, -9.48\%) at NNLO accuracy which is closer to the range (+13.36\%, -7.89\%) for NNLO + $\rm \overline{NNLL}$ level at central rapidity region. This observation along with earlier noticed improvement in perturbative convergence and reliability by the inclusion of $\rm \overline{NNLL}$ contribution, makes a strong case for the relevance of SV+NSV resummed predictions. In addition, there is a systematic decrease in the scale uncertainty when we go from LO + $\rm \overline{LL}$ to NNLO + $\rm \overline{NNLL}$ accuracy. Also, the higher order uncertainty bands are completely included within the lower order uncertainty bands for the resummed predictions. This again indicates that the inclusion of resummed result makes the perturbative expansion of the rapidity distribution more convergent. Table \ref{tab:FONSV7ptmZ} provides the fixed order as well as SV+NSV resummed results at the central scale $\mu_R = \mu_F = m_H/2$ for different benchmark rapidity values at various perturbative orders. It also gives the maximum increments and decrements for the corresponding fixed order and resummed results in the rapidity distribution from the value at the central scale by varying $\{ \mu_R , \mu_F \}$ in the range $\{1/4,1 \} m_H$. 

The above analysis presents compelling arguments to establish the significance of SV+NSV resummed contributions. Now, in order to understand the behaviour of the resummed result w.r.t $\mu_R$ and $\mu_F$ scales in a better way, we study the impact of each scale individually by keeping the other fixed. 
\begin{table*}
 \renewcommand{\arraystretch}{1.9}
\begin{tabular}{ |P{1.0cm}||P{1.8cm}|P{1.8cm}|P{2.0cm}|P{1.9cm} |P{2.0cm}|P{2.0cm}|}
 \hline
  y &$\rm{K_{LO+\rm{LL}}}$&$\rm{K_{LO+\rm{\overline{LL}}}}$&$\rm{K_{NLO+\rm{NLL}}}$&$\rm{K_{NLO+\rm{\overline{NLL}}}}$&$\rm{K_{NNLO+\rm{NNLL}}}$&$\rm{K_{NNLO+\rm{\overline{NNLL}}}}$\\
 \hline
\hline
 0-0.4   & 1.453 & 1.775 & 1.730 & 1.854 & 2.04 & 2.030 \\
 0.4-0.8 & 1.455 & 1.776 & 1.704 & 1.828 & 2.028 & 2.019  \\
 0.8-1.2 & 1.471 & 1.796 & 1.679 & 1.804 & 1.981 & 1.973 \\
 1.2-1.6 & 1.484 & 1.812 & 1.636 & 1.763 & 1.910 & 1.904 \\
 1.6-2.0 & 1.518 & 1.853 & 1.581 & 1.711 & 1.851 & 1.849 \\
 2.0-2.4 & 1.546 & 1.891 & 1.508 & 1.640 & 1.751 & 1.752 \\
\hline
\end{tabular}
\caption{K-factor values of fixed order and resummed results at the central scale $\mu_R=\mu_F= m_H/2$.} \label{tab:SV-NSV}
\end{table*}
\subsection{\textit{Uncertainties of the resummed result with respect to $\mu_R$ and $\mu_F$}}
Here, we demonstrate the individual effect of renormalisation and factorisation scales on the fixed order and resummed results by keeping one of the scales fixed. We start with Fig.\ref{fig:muFfixednsv} where the fixed order(left panel) and SV+NSV resummed(right panel) results for the rapidity distribution are plotted as a function of the rapidity $y$ while keeping the factorisation scale fixed at $\mu_F = m_H/2$. The bands are obtained by varying the renormalisation scale $\mu_R$ in the range $\{ 1/4 , 1 \} m_H$ around the central scale. We find that from NLO accuracy onwards, the addition of resummed contribution to the rapidity distribution decreases the $\mu_R$ dependency of the result. For instance, the uncertainty due to $\mu_R$ scale lies in the range (+23.54\%, -17.36\%) for NLO which reduces to (+17.96\%, -14.91\%) for NLO+$\rm \overline{NLL}$ accuracy. Likewise, the uncertainty ranges between (+7.56\%, -7.89\%) for NNLO +$\rm \overline{NNLL}$ which is an improvement over (+9.48\%, -9.48\%) for NNLO order. Additionally, we observe that there is a considerable reduction in the $\mu_R$ uncertainty while going from LO+$\rm \overline{LL}$ to NNLO +$\rm \overline{NNLL}$ and the higher order uncertainty bands are included within lower order uncertainty bands. The above observations are different from what we had seen in Fig (\ref{fig:7ptfixedsv}) for the 7-point scale variation where the combined uncertainty bands due to $\mu_R$ and $\mu_F$ were wider for resummed predictions. This suggests that SV+NSV resummed results have considerable dependence on factorisation scale $\mu_F$. To explore these observations in detail, we next study the sensitivity of resummed predictions on factorisation scale. 

Fig.\ref{fig:muRfixednsv} depicts the variation of the fixed order and SV+NSV resummed predictions with respect to the factorisation($\mu_F$) scale. The bin-integrated rapidity distribution at various perturbative orders have been plotted against the rapidity $y$ keeping the renormalisation scale fixed at $\mu_R = m_H/2$. The uncertainty bands are obtained by varying $\mu_F$ in the range $\{ 1/4,1 \} m_H$ around the central scale $\mu_R = \mu_F = m_H/2$. We first look at the behaviour of fixed order results(left panel). Interestingly, we see that the uncertainty bands for the fixed order predictions are very thin indicating negligible dependence on the factorisation scale $\mu_F$. This feature is also evident when we look at the 7-point scale varitaion(Fig.\ref{fig:7ptfixedsv}) and $\mu_R$ scale variation(Fig.\ref{fig:muFfixednsv}) plots of fixed order results. The similarity between these two plots indicates that the width of the uncertainty band in Fig.\ref{fig:7ptfixedsv} is mainly due to the variations in the $\mu_R$ scale. On the contrary, the plots for the resummed rapidity distribution(right panel) shows significant dependence on $\mu_F$ scale. The uncertainty due to $\mu_F$ variation lies between (+6.42\%, -8.93\%) for LO+$\rm \overline{LL}$ which escalates to (+32\%, -18.48\%) for NLO+$\rm \overline{NLL}$ accuracy for central rapidity region. It comes down to (+13.36\%, -6.22\%) for NNLO+$\rm \overline{NNLL}$ level around $y=0$. If we compare the 7-point variation plot(Fig.\ref{fig:7ptfixedsv}) and the $\mu_F$ variation plot(Fig.\ref{fig:muRfixednsv}) for resummed predictions of the rapidity distribution, we find that from NLO+ $\rm \overline{NLL}$ level onwards, the scale uncertainty is mainly driven by the variation in $\mu_F$.

To summarise, we did a comparative study between fixed order and SV+NSV resummed predictions for the rapidity distribution. Through the K-factor analysis, we find that the inclusion of resummed results enhances the fixed order predictions till NLO accuracy. At NNLO level, the contributions coming from $\rm \overline{NNLL}$ reduces the fixed order NNLO prediction leading to better perturbative convergence and a more reliable result. The study of the behaviour of resummed results w.r.t the variations in $\mu_R$ scale showed that there is a substantial reduction in the scale dependency as compared to the fixed order results. However, with respect to factorisation scale, the fixed order results show negligible dependence while the addition of resummed contributions increases the sensitivity of the rapidity distribution. The $\mu_F$ scale variation is more at NLO+$\rm \overline{NLL}$ level as compared to NNLO+$\rm \overline{NNLL}$ accuracy. This behaviour of the resummed result w.r.t $\mu_F$ variation could be naively attributed to the absence of NSV logarithms coming from the off-diagonal channel. We know that under $\mu_F$ scale variation, the partonic channels get mixed due to DGLAP evolution of the parton distribution functions and compensates among each other to reduce the $\mu_F$ dependency. Hence, the lack of off-diagonal resummed NSV logarithms could be the reason for the increased sensitivity of the SV+NSV resummed result. This is not correct because for Higgs Boson production through gluon fusion, the off-diagonal $qg$ channel has minuscule contribution. To understand the reason behind this, we study the behaviour of SV resummed result in comparison to SV+NSV resummed result in the next section.

\subsection{SV vs SV+NSV resummed rapidity distribution}

In this section, we will investigate the impact of inclusion of resummed NSV logarithms by comparing it with the well established SV resummed results. The analysis in our previous section showed that $\mu_R$ scale uncertainty gets improved by adding SV+NSV resummed contribution to the fixed order result. In contrary to this, there is a significant increase in the $\mu_F$ scale uncertainty by the inclusion of resummed results. Here, we try to understand which part of SV+NSV resummed result is responsible for the exhibited behaviour w.r.t $\mu_R$ and $\mu_F$ scales.
\begin{figure*}
\centering
\includegraphics[scale=0.40]{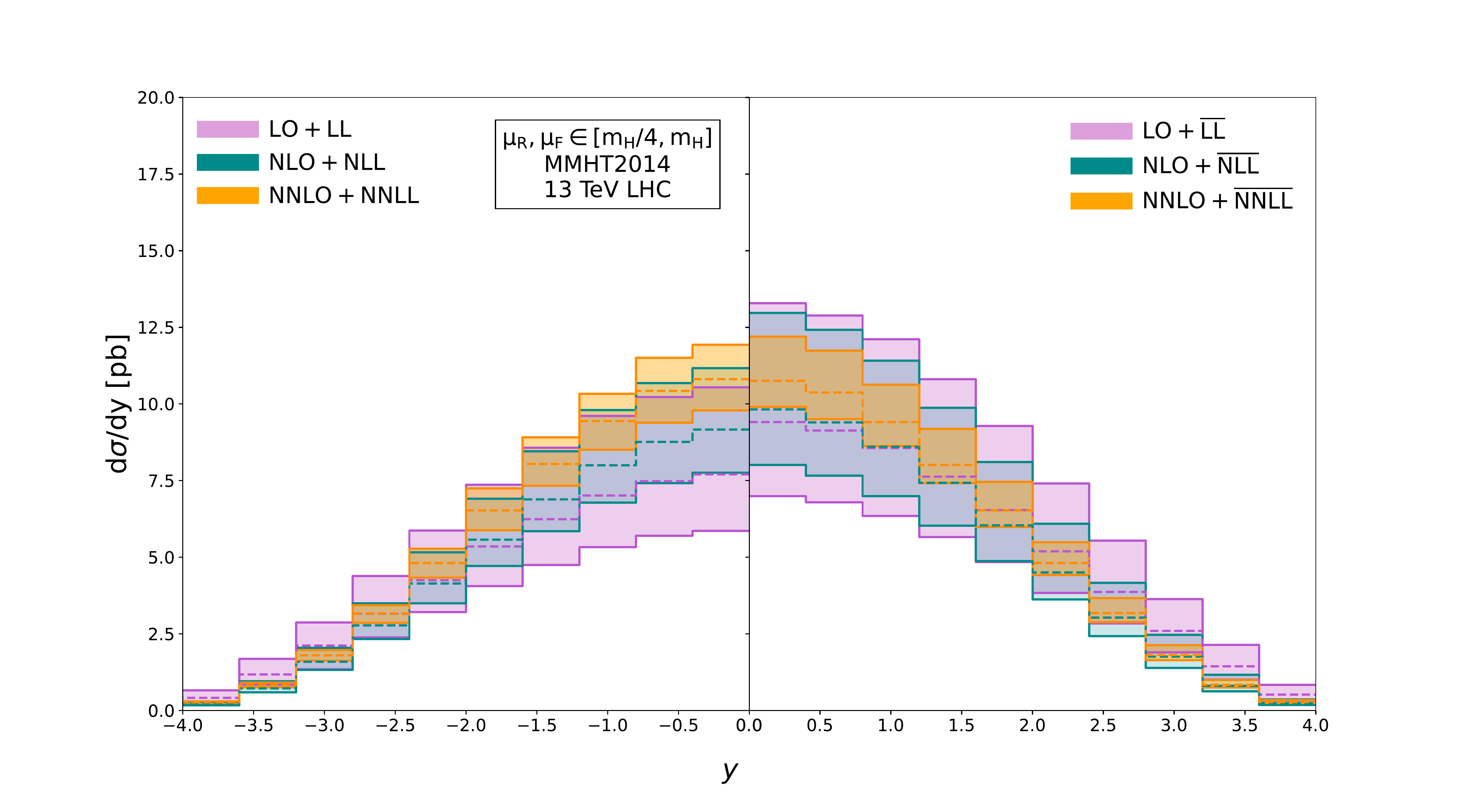}
\caption{Comparison of 7-point scale variation between SV and SV+NSV resummed results for 13 TeV LHC. The dashed lines refer to the corresponding central scale
values at each order.}
\label{fig:7ptSVNSV}
\end{figure*}

\begin{table*}
 \renewcommand{\arraystretch}{2.7}
\begin{tabular}{ |P{1.0cm}||P{2.2cm}|P{2.2cm}|P{2.2cm}||P{2.2cm}|P{2.2cm}|P{2.2cm}|}
 \hline
  y &LO+$\rm{LL}$&LO+$\rm{\overline{LL}}$&NLO+$\rm{NLL}$&NLO+$\rm{\overline{NLL}}$&NNLO+$\rm{NNLL}$&NNLO+$\rm{\overline{NNLL}}$\\
 \hline
\hline
 0-0.4 & 7.6990 $^{+ 2.840 }_{- 1.835 }$ 
 & 9.406 $^{+ 3.878 }_{- 2.411 }$ 
 & 9.1660 $^{+ 2.00 }_{- 1.410 }$ 
 & 9.8252 $^{+ 3.144 }_{- 1.816 }$ 
 & 10.8117 $^{+ 1.124 }_{- 1.025 }$ 
 & 10.7559 $^{+ 1.437 }_{- 0.849 }$  \\
 
 0.4-0.8 & 7.4784 $^{+ 2.747 }_{- 1.780 }$ 
 & 9.1303 $^{+ 3.752 }_{- 2.338 }$ 
 & 8.7608 $^{+ 1.922 }_{- 1.343 }$ 
 & 9.3991 $^{+ 3.023 }_{- 1.739 }$ 
 & 10.4274 $^{+ 1.079 }_{- 1.041 }$ 
 & 10.3777 $^{+ 1.361 }_{- 0.872 }$ \\

 0.8-1.2 & 7.0128 $^{+ 2.599 }_{- 1.687 }$ 
 & 8.561 $^{+ 3.548 }_{- 2.211 }$ 
 & 8.0011 $^{+ 1.795 }_{- 1.217 }$ 
 & 8.601 $^{+ 2.815 }_{- 1.608 }$ 
 & 9.4445 $^{+ 0.889 }_{- 0.940 }$ 
 & 9.407 $^{+ 1.222 }_{- 0.783 }$ \\

 1.2-1.6 & 6.2455 $^{+ 2.331 }_{- 1.499 }$ 
 & 7.626 $^{+ 3.179 }_{- 1.967 }$ 
 & 6.8874 $^{+ 1.564 }_{- 1.037 }$ 
 & 7.4229 $^{+ 2.457 }_{- 1.393 }$ 
 & 8.0385 $^{+ 0.869 }_{- 0.708 }$ 
 & 8.0143 $^{+ 1.170 }_{- 0.594 }$  \\
 
 1.6-2.0 & 5.3542 $^{+ 2.012 }_{- 1.296 }$ 
 & 6.537 $^{+ 2.749 }_{- 1.70  }$ 
 & 5.577 $^{+ 1.327 }_{- 0.858 }$ 
 & 6.036 $^{+ 2.074 }_{- 1.159 }$ 
 & 6.5317 $^{+ 0.716 }_{- 0.652 }$ 
 & 6.524 $^{+ 0.937 }_{- 0.537 }$ \\
 
 2.0-2.4 & 4.2462 $^{+ 1.622 }_{- 1.034 }$ 
 & 5.1918 $^{+ 2.213 }_{- 1.358 }$ 
 & 4.1411 $^{+ 1.019 }_{- 0.644 }$ 
 & 4.5031 $^{+ 1.592 }_{- 0.878 }$ 
 & 4.8088 $^{+ 0.467 }_{- 0.478 }$ 
 & 4.8123 $^{+ 0.677 }_{- 0.390 }$  \\
\hline
\end{tabular}
\caption{Values of resummed rapidity distribution at various orders in comparison to the fixed order results in pb at the central scale $\mu_R=\mu_F= m_H/2$ for 13 TeV LHC.} \label{tab:SVNSV7pt}
\end{table*}

We first look at the K-factor values given in Table \ref{tab:SV-NSV} for SV and SV+NSV resummed results at various perturbative orders for benchmark rapidity values. We find that the inclusion of resummed NSV contributions enhances the rapidity distribution by 7.17\% when we go from $\rm NLL$ to $\rm \overline{NLL}$ accuracy at the central rapidity region. On the other hand, there is a slight reduction of 0.49\% in the rapidity distribution when we go from $\rm NNLL$ to $\rm \overline{NNLL}$ accuracy. We also observe that the K-factor values for NLO+$\rm \overline{NLL}$ and NNLO+$\rm \overline{NNLL}$ are closer to each other as compared to the corresponding values at NLO+$\rm NLL$ and NNLO+$\rm NNLL$. The above observations suggest that the incorporation of resummed NSV contribution to the threshold SV resummed result improves the perturbative convergence of the predictions.

We move on to investigate the uncertainties related to $\mu_R$ and $\mu_F$ scales arising from the NSV logarithms. We start with the canonical 7-point scale variation plot shown in Fig.\ref{fig:7ptSVNSV} for the bin-integrated rapidity distribution of the Higgs boson for SV(left panel) and SV+NSV(right panel) resummed predictions at various perturbative orders. The scales $\{ \mu_R, \mu_F \}$ is varied in the range $\{ m_H/4, m_H \}$,  keeping the ratio $\frac{\mu_R}{\mu_F}$ not larger than 2 and smaller than 1/2 around the central scale value $\mu_R = \mu_F = m_H/2$ for 13 TeV LHC. From Fig.\ref{fig:7ptSVNSV}, we notice that the addition of resummed NSV contributions to the SV resummed results increase the combined scale uncertainty till next-to-next-to leading logarithmic accuracy. However, the difference in the width of uncertainty bands of NNLO+$\rm \overline{NNLL}$ and NNLO+$\rm NNLL$ is significantly less as compared to that of NLO+$\rm \overline{NLL}$ and NLO+$\rm NLL$. For instance, the 7-point scale uncertainty at NLO+$\rm NLL$ varies between (+21.82\%, -15.38\%) which is considerably increased to (+32\%, -18.48\%) at NLO+$\rm \overline{NLL}$ around central rapidity region. But, the uncertainty of (+13.36\%, -7.89\%) at NNLO+$\rm \overline{NNLL}$ is relatively not too larger than the uncertainty of (+10.4\%, -9.48\%) at NNLO+$\rm NNLL$ accuracy. We can also see that the uncertainty bands of higher order SV+NSV resummed results are completely within the lower order ones over the full rapidity region which is not the case with SV resummed predictions. This hints towards a more convergent perturbative expansion by the incorporation of resummed NSV logarithms. In Table \ref{tab:SVNSV7pt}, we have given the rapidity distributions of SV and SV+NSV resummed predictions at central scale $\mu_R = \mu_F = m_H/2$ for benchmark rapidity values along with the corresponding maximum increments and decrements in the rapidity distribution. The increments and decrements from the central scale values are calculated by varying $\{ \mu_R , \mu_F \}$ in the range $\{ 1/4 , 1 \} m_H $ . The above given percentage uncertainties for various perturbative orders are calculated using values from Table \ref{tab:SVNSV7pt}.

In the above paragraph, we compared the SV and SV+NSV resummed results for the 7-point scale variations. We found that the resummed NSV logarithms spoil the combined scale uncertainty especially at NLO+$\rm \overline{NLL}$ accuracy. Let us now turn to compare the SV and SV+NSV resummed predictions under the variation of each of these scales individually and try to reason out the behaviour exhibited by the resummed NSV logarithms.
\begin{figure*}
\includegraphics[scale=0.40]{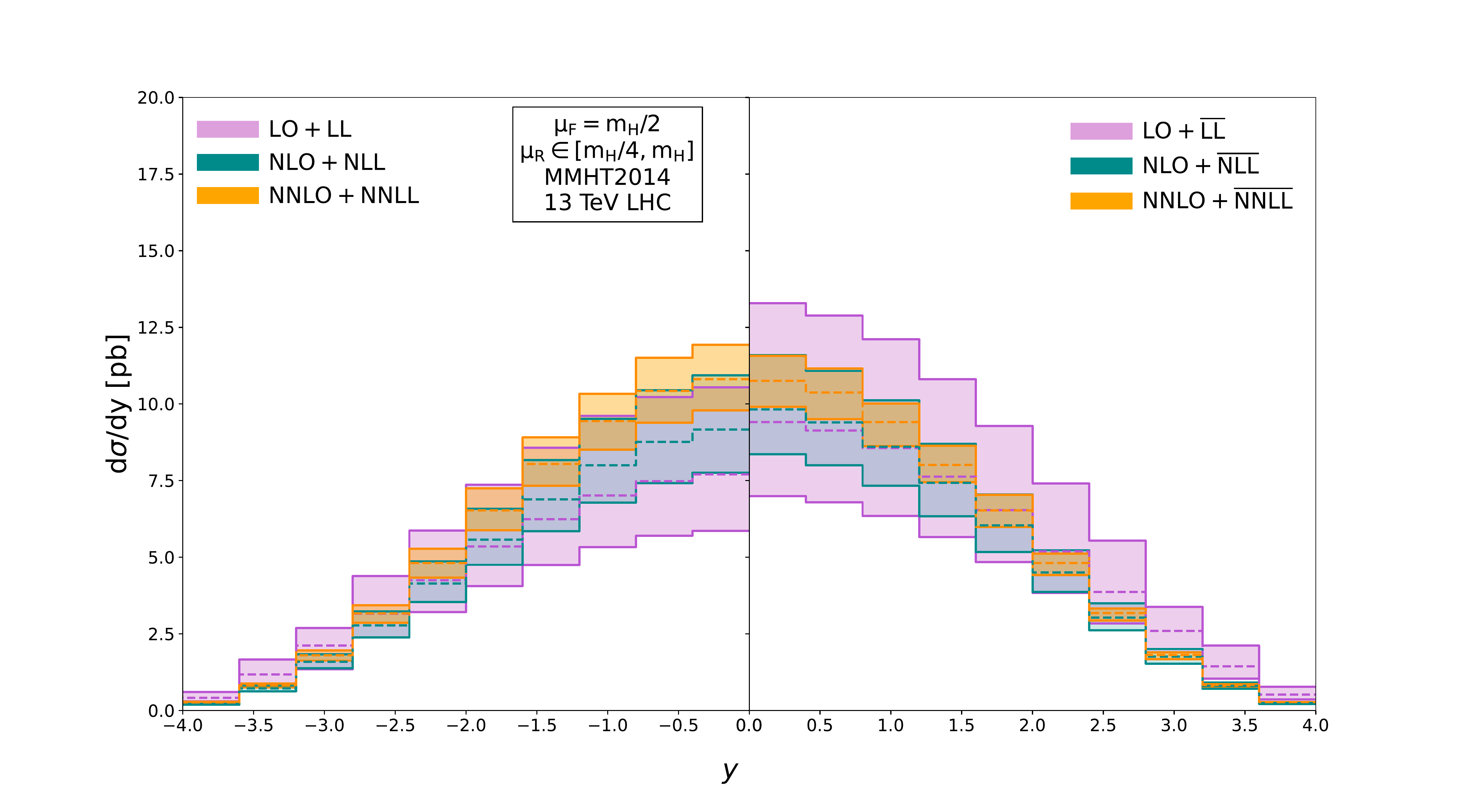}
\caption{Comparison of $\mu_R$ scale variation between SV and SV+NSV resummed results with the scale $\mu_F = m_H/2$. The dashed lines refer to the corresponding central scale
values at each order.}
\label{fig:muR2Tsvnsv}
\end{figure*}

\begin{figure*}
\includegraphics[scale=0.40]{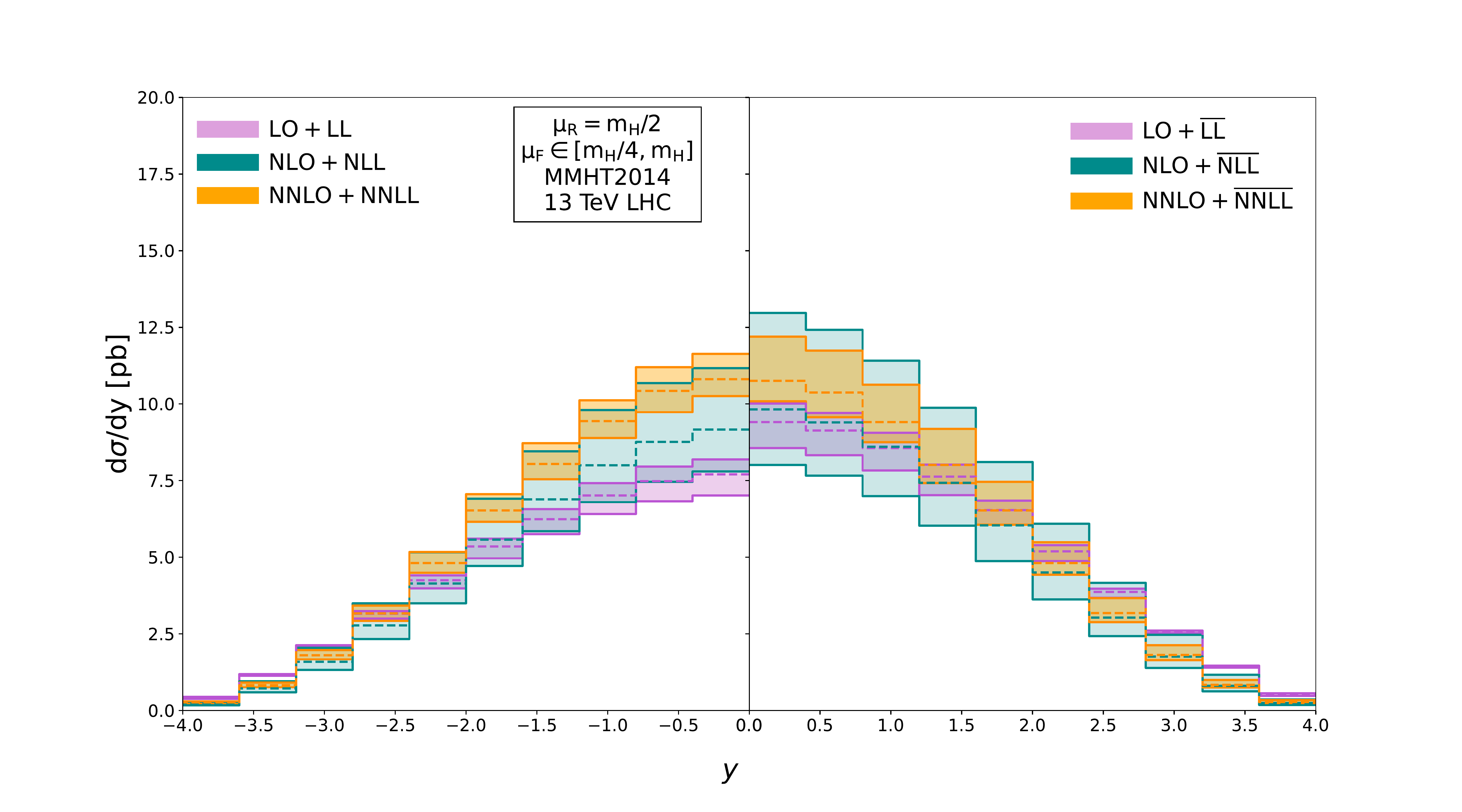}
\caption{Comparison of $\mu_F$ scale variation between SV and SV+NSV resummed results with the scale $\mu_R = m_H/2$. The dashed lines refer to the corresponding central scale
values at each order.}
\label{fig:muF2Tsvnsv}
\end{figure*}

We first compare $\mu_R$ scale uncertainties of SV and SV+NSV resummed predictions. Fig \ref{fig:muR2Tsvnsv} illustrates the bin integrated rapidity distributions as a function of $y$ under $\mu_R$ scale variation with $\mu_F = m_H/2$ kept fixed for both SV (left panel) and SV+NSV (right panel) resummed results. The bands are obtained by varying 
$\{ \mu_R , \mu_F \}$ in the range $\{ 1/4 , 1 \} m_H $ around the central scale $\mu_R = \mu_F = m_H/2$ for 13 TeV LHC. From Fig \ref{fig:muR2Tsvnsv}, we can see that the uncertainty bands for SV+NSV resummed predictions become narrower as compared to the SV resummed result from next-to-leading order onwards. Quantitatively, the $\mu_R$ scale uncertainty for NLO+$\rm \overline{NLL}$ varying between (+17.96\%, -14.91\% ) is less than the corresponding uncertainty of (+19.35\%, -15.39\% ) for NLO+$\rm NLL$ accuracy around $y=0$. Similarly, the uncertainty of (+7.56\%, -7.89\% ) at NNLO+$\rm \overline{NNLL}$ due to $\mu_R$ scale variation is a considerable reduction from the uncertainty lying between (+10.40\%, -9.48\% ) at NNLO+$\rm NNLL$ accuracy. As mentioned earlier, the $\mu_R$ scale uncertainty for fixed order results lie in the range (+23.54\%, -17.36\% ) at NLO and for NNLO order it varies between (+9.48\%, -9.48\% ) around the central rapidity region. These percentages show that at NLO level, the addition of resummed SV as well as resummed NSV contributions to the fixed order rapidity distribution improve the renormalisation scale uncertainty. This is expected as the inclusion of higher order logarithmic corrections within a particular channel leads to decrease in sensitivity of the rapidity distribution w.r.t $\mu_R$ scale. Nevertheless, the addition of resummed SV contributions to the fixed order result at NNLO level does not bring any notable change in $\mu_R$ uncertainty. The SV distributions constitute only 15.81\% of the Born cross section whereas NSV logarithms contribute to overall 58.91\% of the Born cross section at NNLO for the case of inclusive cross section of Higgs boson production as shown in Table 2 of Ref.\cite{Ajjath:2021bbm}. The same trend is expected to follow for the rapidity distribution of Higgs Boson production through gluon fusion as well. Thus, SV contributions being the sub-dominant contributor is not able to bring any change in the behaviour of fixed order result w.r.t $\mu_R$ scale variation at NNLO. On the other hand, inclusion of NSV logarithms which is the dominant contributor at this order results in significant improvement in the $\mu_R$ scale uncertainty of the rapidity distribution. These observations are evident from the uncertainty percentages given above.

Based on the inferences given above, it is established that the resummed results improve the renormalisation scale uncertainty of the rapidity distribution. We would like to mention that the resummed results not only carry the all-order information of the distributions and logarithms that we are resumming, but they also contain certain spurious terms resulting from the "inexact" Mellin inversion of the $N$-space resummed result. These spurious terms are beyond the precision of the resummed quantity. For instance, the resummed SV result at NNLO+$\rm NNLL$ accuracy contains the all-order correction arising from the summation of next-to-next-to leading towers of SV distributions and in addition certain spurious NSV and beyond NSV terms. Similarly, the resummed NSV logarithms at this order contains the all-order correction arising from the summation of next-to-next-to leading towers of NSV logarithms and certain spurious beyond NSV terms. These spurious terms play an important role in the factorisation scale variation of the resummed SV+NSV predictions. 

Next, we show the comparison of SV and SV+NSV resummed results for the rapidity distribution under $\mu_F$ scale variation. We plot SV(left panel) and SV+NSV(right panel) resummed bin integrated rapidity distributions as a function of $y$ keeping the renormalisation scale fixed at $\mu_R = m_H/2$ as depicted in Fig \ref{fig:muF2Tsvnsv}. The bands are obtained by varying $\{ \mu_R , \mu_F \}$ in the range $\{ 1/4 , 1 \} m_H $ around the central scale $\mu_R = \mu_F = m_H/2$ for 13 TeV LHC. We observe that the uncertainty bands of SV+NSV resummed results are wider than the corresponding bands of SV resummed predictions at every order till NNLO+$\rm \overline{NNLL}$. This can be seen quantitatively from the $\mu_F$ scale uncertainty of (+21.82\%, -14.89\%) for NLO+$\rm NLL$ which gets escalated to (+32.0\%, -18.48\%) for NLO+$\rm \overline{NLL}$ around the central rapidity region. In the same way, the uncertainty lies between (+13.36\%, -6.22\%) for NNLO+$\rm \overline{NNLL}$ which is a considerable increment over (+7.63\%, -5.14\%) for NNLO+$\rm NNLL$ around $y=0$. 

Let us examine more closely the reason behind the significant dependence of SV as well as SV+NSV resummed results on $\mu_F$ scale variation. We consider Fig.(4) given in Ref.\cite{Ajjath:2021bbm} for this purpose. We have plotted the fixed order results truncated to SV+NSV accuracy for the inclusive cross section against $\mu_F$ scale variation. We observe that there is a significant dependence of the truncated result on the $\mu_F$ scale variation from NLO level onwards which escalates at NNLO accuracy. The same behaviour is expected to be manifested by the fixed order results truncated to SV+NSV accuracy for the rapidity distribution as well. Therefore, we compare the behaviour of this plot with that of the full fixed order results under $\mu_F$ scale variation depicted in the left panel of Fig.\ref{fig:muRfixednsv}. We find that the $\mu_F$ scale dependence of the fixed order results of the rapidity distribution shown in Fig.\ref{fig:muRfixednsv} is very mild. This indicates that the large $\mu_F$ scale variation exhibited by fixed order results truncated to SV+NSV accuracy is expected to get compensation from beyond NSV terms in the threshold expansion. Fig.(4) given in the Ref.\cite{Ajjath:2021bbm} also suggests that the contributions coming from the beyond NSV terms become more significant with the increase in the order of perturbative series. It is because more compensation is required to make the fixed order predictions almost independent of the $\mu_F$ scale variation as we go to higher orders in the perturbative expansion.

Now, we explore the reason behind the behaviour of resummed SV results under $\mu_F$ scale variation. The addition of resummed SV contributions to the fixed order predictions increase the sensitivity of the rapidity distribution w.r.t $\mu_F$ scale variation. This can be seen quantitatively from $\mu_F$ scale uncertainties varying in the range (+21.82\%, -14.89\%) and (+7.63\%, -5.14\%) for NLO + $\rm NLL$ and NNLO + $\rm NNLL$ respectively around the central rapidity region. The spurious beyond SV terms resulting from the "inexact" Mellin inversion of the $N$-space resummed result is mainly responsible for this uncertainty. We also observe that the $\mu_F$ scale uncertainty decreases significantly at NNLO + $\rm NNLL$ as compared to NLO + $\rm NLL$ accuracy. This suggests that the resummation of next-to-next-to leading SV distributions compensates the uncertainty arising from the spurious terms of NLO + $\rm NLL$ with the higher order logarithmic corrections.

We now proceed to analyse the behaviour of NSV logarithmns w.r.t $\mu_F$ scale variation using the observations made in the above paragraphs for fixed order and SV resummed results for the rapidity distributions. From Fig.(4) of Ref.\cite{Ajjath:2021bbm} and the behaviour of fixed order results under $\mu_F$ variation shown in Fig.\ref{fig:muRfixednsv}, we deduced that the $\mu_F$ uncertainties due to SV+NSV terms are compensated by uncertainties arising from beyond NSV terms in perturbation theory. We also observed that this compensation increases with the increase in the order of perturbation theory. We know that the SV+NSV resummed predictions contain spurious beyond NSV terms resulting from the resummation of SV as well as NSV parts.
The behaviour of resummed SV result showed that the corresponding spurious terms increase the $\mu_F$ scale variation of the SV resummed rapidity disribution. Now, with the inclusion of NSV resummed logarithms, there is one additional source to generate spurious beyond NSV terms. Thus, the resummation of NSV logarithms is not supposed to improve the uncertainties arising due to $\mu_F$ scale variation of SV resummed result. This inference is consistent with the plots shown in Fig.\ref{fig:muF2Tsvnsv} where the inclusion of resummed NSV logarithms to the SV resummed result increase the uncertainty due to $\mu_F$ scale variation till NNLO+$\rm \overline{NNLL}$ accuracy. We also observe that the uncertainty drops down significantly when we go from NLO+$\rm \overline{NLL}$ to NNLO+$\rm \overline{NNLL}$. This suggests that the higher order logarithmic corrections from the NSV terms improve the $\mu_F$ dependency of SV+NSV resummed results. They do so by adding more terms and also by compensating for the spurious NSV terms arising due to SV resummation at lower logarithmic accuracy.

In summary, the renormalisation scale dependency of the rapidity distribution decreases by the inclusion of resummed NSV logarithms. This is expected because any change resulting from the $\mu_R$ variation get compensated by the addition of higher order terms coming from the resummation of SV and NSV logarithms to all orders. This scenario changes for the case of factorisation scale variation. The fixed order rapidity distribution shows negligible dependence on $\mu_F$ scale variation. The addition of SV resummed terms increases the sensitivity of the result w.r.t $\mu_F$ scale which further deteriorates by the inclusion of resummed NSV logarithms. One of the reason for this behaviour is the presence of spurious terms arising due to the "inexact" Mellin inversion of the $N$-space resumed result. In addition, this significant $\mu_F$ dependency of the SV+NSV resummed rapidity distribution is attributed to the lack of beyond NSV terms. This hints towards the importance of beyond NSV terms to get a more accurate and reliable prediction for rapidity distribution. We would also like to mention that we have used the same PDF set for both fixed order as well as resumed predictions. It is worthwhile to consider resummed PDFs if they are available especially for studying the variations w.r.t $\mu_F$ scale.



\section{Discussion and Conclusion}
In this article, we provide for the first time, the phenomenological predictions for resummed next-to soft virtual corrections to the rapidity distribution of Higgs production in gluon fusion up to NNLO + $\rm \overline{NNLL}$ accuracy.
We have used our recent formalism \cite{Ajjath:2020lwb} to systematically resum NSV logarithms arising from diagonal gluon-gluon ($gg$) channel to all orders. 
In our previous work on inclusive Higgs cross section, we have studied the significance of NSV logarithms in the fixed order predictions \cite{Ajjath:2021bbm}. The interesting results and findings of \cite{Ajjath:2021bbm} have been an inspiration to study the numerical relevance of these NSV logarithms in the case of rapidity distribution as well. 

We have analysed the numerical effects of SV+NSV higher order predictions by providing the K-factor values around the central scale $\mu_R = \mu_F = m_H/2$ for benchmark rapidity values. We find that there is an enhancement of 77.5\% and 4.04\% at LO+$\rm \overline{LL}$ and NLO+$\rm \overline{NLL}$  respectively by the inclusion of SV+NSV resummed results around the central rapidity region. However, the rapidity distribution at NNLO gets decreased by 2.96\% when we include the $\rm \overline{NNLL}$ resummed corrections at the central rapidity region.
This clearly indicates the improvement in perturbative convergence at NNLO+$\rm \overline{NNLL}$ accuracy. Furthermore, we notice that the inclusion of higher order resummed SV+NSV corrections makes the perturbative predictions more reliable due to very small percentage enhancement in the rapidity distribution at NNLO+$\rm \overline{NNLL}$ in comparison to NLO+$\rm \overline{NLL}$ for a wide range of rapidity values.

The standard canonical 7-point scale variation approach has been employed to study the dependence of our numerical predictions on renormalisation ($\mu_R$) and factorisation($\mu_F$) scales. We have presented the plot of 7-point scale variation around the central scale $\mu_R = \mu_F = m_H/2$ for 13 TeV LHC. We find that the width of uncertainty bands of resummed predictions is more than that of the corresponding fixed order results till NLO. However, for NNLO and NNLO+$\rm \overline{NNLL}$, the width of the uncertainty bands are comparable. Thus, by performing the 7-point scale variation analysis, we observe that there is a systematic reduction in the uncertainty of the resummed results while going to higher logarithmic accuracy for the central scale $\mu_R = \mu_F = m_H/2$ around the central rapidity region. Moreover, we notice that the uncertainty bands corresponding to higher order predictions are well contained within that of lower order ones. This is also an indication of better perturbative convergence attained by the process of resummation. 

The detailed analysis of the scale uncertainties unveiled that the 7-point scale uncertainties of SV+NSV resummed predictions are mostly driven by the variations in factorisation scale $\mu_F$ especially at NLO + $\rm \overline{NLL}$. On the other hand, the dependence on the renormalisation scale  $\mu_R$ gets reduced by the inclusion of SV+NSV resummed results leading to more reliable predictions. Furthermore, there is a systematic reduction in the $\mu_R$ scale uncertainty while going from LO + $\rm \overline{LL}$ to NNLO + $\rm \overline{NNLL}$ due to the addition of higher logarithmic corrections.   
From the comparison of SV and SV+NSV resummed results, we find that it is the NSV part of the resummation which is responsible for bringing down the uncertainty due to $\mu_R$ scale variation. Thus, the inclusion of more corrections within the same partonic channel improves the $\mu_R$ scale  uncertainties. This is due to the fact that different channels being renormalisation group invariant, do not mix under $\mu_R$ scale variation. However, the uncertainties due to $\mu_F$ scale variations get worse by the addition of NSV corrections. From our analysis, we found that the lack of beyond NSV resummed terms is the reason behind the sensitivity of SV+NSV resummed results on $\mu_F$ scale variations.   

\section{Acknowledgements}

We thank J. Michel and F. Tackmann for
third order results of rapidity for comparing purposes and 
C. Duhr and B. Mistlberger for providing third order results for the inclusive reactions. In addition we would also like to thank the computer administrative unit of IMSc for their help and support.

\appendix
\begin{widetext}
\section{NSV Resummation exponents $   \overline{g}_{d,i}^g(\omega)$} \label{app:gbdN}
\begin{align}
\begin{autobreak}
   \overline{g}_{d,1}^g(\omega) =

        \frac{1}{\beta_0}C_A   \bigg\{
          2 L_\omega
          \bigg\}\,,
   
\end{autobreak}\\
  \begin{autobreak}
   \overline{g}_{d,2}^g(\omega) =

       \frac{\beta_1}{\beta_0^2} C_A   \bigg\{
           2 \omega
          + 2 L_\omega
          \bigg\}

       +\frac{1}{\beta_0}C_A n_f   \bigg\{
           \frac{20}{9} \omega
          \bigg\}

       +\frac{1}{\beta_0}C_A^2   \bigg\{
          - \frac{134}{9} \omega
          + 4 \omega \zeta_2
          \bigg\}

       + C_A   \bigg\{
          - 2
          + 2 L_{qr}
          - 2 L_{fr}
          + 2 L_{fr} \omega
          - 4 \gamma_E
          \bigg\}\,,
\end{autobreak}\\
  \begin{autobreak}

   \overline{g}_{d,3}^g(\omega) =

       \frac{\beta_1^2}{\beta_0^3} C_A   \bigg\{
          \omega^2
          - L_\omega^2
          \bigg\}

       + \frac{\beta_2}{\beta_0^2} C_A   \bigg\{
          - \omega^2
          \bigg\}

       + \frac{\beta_1}{\beta_0^2} C_A n_f   \bigg\{
          - \frac{20}{9} \omega
          + \frac{10}{9} \omega^2
          - \frac{20}{9} L_\omega
          \bigg\}

       + \frac{\beta_1}{\beta_0^2} C_A^2   \bigg\{
           \frac{134}{9} \omega
          - 4 \omega \zeta_2
          - \frac{67}{9} \omega^2
          + 2 \omega^2 \zeta_2
          + \frac{134}{9} L_\omega
          - 4 L_\omega \zeta_2
          \bigg\}

       +\frac{1}{\beta_0}C_A n_f^2   \bigg\{
           \frac{8}{27} \omega
          - \frac{4}{27} \omega^2
          \bigg\}

       +\frac{1}{\beta_0}C_A C_F n_f   \bigg\{
           \frac{55}{3} \omega
          - 16 \omega \zeta_3
          - \frac{55}{6} \omega^2
          + 8 \omega^2 \zeta_3
          \bigg\}

       +\frac{1}{\beta_0}C_A^2 n_f   \bigg\{
           \frac{418}{27} \omega
          + \frac{56}{3} \omega \zeta_3
          - \frac{80}{9} \omega \zeta_2
          - \frac{209}{27} \omega^2
          - \frac{28}{3} \omega^2 \zeta_3
          + \frac{40}{9} \omega^2 \zeta_2
          \bigg\}

       +\frac{1}{\beta_0}C_A^3   \bigg\{
          - \frac{245}{3} \omega
          - \frac{44}{3} \omega \zeta_3
          + \frac{536}{9} \omega \zeta_2
          - \frac{88}{5} \omega \zeta_2^2
          + \frac{245}{6} \omega^2
          + \frac{22}{3} \omega^2 \zeta_3
          - \frac{268}{9} \omega^2 \zeta_2
          + \frac{44}{5} \omega^2 \zeta_2^2
          \bigg\}

       + \frac{\beta_1}{\beta_0} C_A   \bigg\{
           2 L_\omega
          - 2 L_\omega L_{qr}
          + 4 L_\omega \gamma_E
          \bigg\}

       + C_A n_f   \bigg\{
           \frac{116}{27}
          - \frac{2}{3} \zeta_2
          - \frac{20}{9} L_{qr}
          + \frac{20}{9} L_{fr}
          - \frac{40}{9} L_{fr} \omega
          + \frac{20}{9} L_{fr} \omega^2
          + \frac{40}{9} \gamma_E
          \bigg\}

       + C_A^2   \bigg\{
          - \frac{806}{27}
          + 14 \zeta_3
          + \frac{23}{3} \zeta_2
          + \frac{134}{9} L_{qr}
          - 4 L_{qr} \zeta_2
          - \frac{134}{9} L_{fr}
          + 4 L_{fr} \zeta_2
          + \frac{268}{9} L_{fr} \omega
          - 8 L_{fr} \omega \zeta_2
          - \frac{134}{9} L_{fr} \omega^2
          + 4 L_{fr} \omega^2 \zeta_2
          - \frac{268}{9} \gamma_E
          + 8 \gamma_E \zeta_2
          \bigg\}

       + \beta_0 C_A   \bigg\{
          - \zeta_2
          + 2 L_{qr}
          - L_{qr}^2
          + L_{fr}^2
          - 2 L_{fr}^2 \omega
          + L_{fr}^2 \omega^2
          - 4 \gamma_E
          + 4 \gamma_E L_{qr}
          - 4 \gamma_E^2
          \bigg\}\,,

\end{autobreak}
\end{align}
\end{widetext}
\begin{widetext}
\section{NSV Resummation exponents $  h^g_{d,ij}(\omega)$} \label{app:hdN}
\begin{align}
\begin{autobreak}  
   h^g_{d,00}(\omega) =

        \frac{1}{\beta_0}  C_A   \bigg\{
          - 4 L_\omega
          \bigg\} \quad 
   
   h^g_{d,01}(\omega) = 0\,,
   
\end{autobreak}\\ 
  \begin{autobreak}
   h^g_{d,10}(\omega) =

       \frac{1}{2 \beta_0^2 (\omega -1)} \Bigg[    \beta_1 C_A   \bigg\{
           8 \omega
          + 8 L_\omega
          \bigg\}

       + \beta_0 C_A n_f   \bigg\{
           \frac{80}{9} \omega
          \bigg\}

       + \beta_0 C_A^2   \bigg\{
          - \frac{536}{9} \omega
          + 16 \omega \zeta_2
          - 32 \gamma_E \omega
          \bigg\}

       + \beta_0^2 C_A   \bigg\{
          - 4
          - 8 L_{fr}
          + 8 L_{fr} \omega
          + 8 L_{qr}
          - 16 \gamma_E 
          \bigg\}\Bigg]\,,

 \end{autobreak}\\ 
  \begin{autobreak}
   h^g_{d,11}(\omega) =

        \frac{ C_A^2}{2 \beta_0 (\omega -1)^2}   \bigg\{
           28 \omega
          - 32 \omega^2
          \bigg\}\,,

  \end{autobreak}\\ 
  \begin{autobreak}
   h^g_{d,20}(\omega) =

      \frac{1}{2 \beta_0^3 (\omega -1)^2} \Bigg[     \beta_1^2 C_A   \bigg\{
          - 4 \omega^2
          + 4 L_\omega^2
          \bigg\}

       + \beta_0 \beta_2 C_A   \bigg\{
           4 \omega^2
          \bigg\}

       + \beta_0 \beta_1 C_A n_f   \bigg\{
           \frac{80}{9} \omega
          - \frac{40}{9} \omega^2
          + \frac{80}{9} L_\omega
          \bigg\}

       + \beta_0 \beta_1 C_A^2   \bigg\{
          - \frac{536}{9} \omega
          + 16 \omega \zeta_2
          + \frac{268}{9} \omega^2
          - 8 \omega^2 \zeta_2
          - 32 \gamma_E \omega
          + 16 \gamma_E \omega^2
          - \frac{536}{9} L_\omega
          + 16 L_\omega \zeta_2
          - 32 L_\omega \gamma_E
          \bigg\}

       + \beta_0^2 C_A n_f^2   \bigg\{
          - \frac{32}{27} \omega
          + \frac{16}{27} \omega^2
          \bigg\}

       + \beta_0^2 C_A C_F n_f   \bigg\{
          - \frac{172}{3} \omega
          + 64 \omega \zeta_3
          + \frac{86}{3} \omega^2
          - 32 \omega^2 \zeta_3
          \bigg\}

       + \beta_0^2 C_A^2 n_f   \bigg\{
          - \frac{1096}{27} \omega
          - \frac{224}{3} \omega \zeta_3
          + \frac{320}{9} \omega \zeta_2
          + \frac{548}{27} \omega^2
          + \frac{112}{3} \omega^2 \zeta_3
          - \frac{160}{9} \omega^2 \zeta_2
          - \frac{640}{9} \gamma_E \omega
          + \frac{320}{9} \gamma_E \omega^2
          \bigg\}

       + \beta_0^2 C_A^3   \bigg\{
           \frac{724}{3} \omega
          - \frac{112}{3} \omega \zeta_3
          - \frac{2144}{9} \omega \zeta_2
          + \frac{352}{5} \omega \zeta_2^2
          - \frac{362}{3} \omega^2
          + \frac{56}{3} \omega^2 \zeta_3
          + \frac{1072}{9} \omega^2 \zeta_2
          - \frac{176}{5} \omega^2 \zeta_2^2
          + \frac{4288}{9} \gamma_E \omega
          - 128 \gamma_E \omega \zeta_2
          - \frac{2144}{9} \gamma_E \omega^2
          + 64 \gamma_E \omega^2 \zeta_2
          \bigg\}

       + \beta_0^2 \beta_1 C_A   \bigg\{
           8 \omega
          - 4 \omega^2
          - 4 L_\omega
          + 8 L_\omega L_{qr}
          - 16 L_\omega \gamma_E
          \bigg\}

       + \beta_0^3 C_A n_f   \bigg\{
          - \frac{272}{27}
          + \frac{32}{3} \zeta_2
          - \frac{80}{9} L_{fr}
          + \frac{160}{9} L_{fr} \omega
          - \frac{80}{9} L_{fr} \omega^2
          + \frac{80}{9} L_{qr}
          - \frac{148}{9} \gamma_E
          \bigg\}

       + \beta_0^3 C_A^2   \bigg\{
           \frac{1808}{27}
          - 56 \zeta_3
          - \frac{224}{3} \zeta_2
          + \frac{536}{9} L_{fr}
          - 16 L_{fr} \zeta_2
          - \frac{1072}{9} L_{fr} \omega
          + 32 L_{fr} \omega \zeta_2
          + \frac{536}{9} L_{fr} \omega^2
          - 16 L_{fr} \omega^2 \zeta_2
          - \frac{536}{9} L_{qr}
          + 16 L_{qr} \zeta_2
          + \frac{1060}{9} \gamma_E
          - 32 \gamma_E \zeta_2
          + 32 \gamma_E L_{fr}
          - 64 \gamma_E L_{fr} \omega
          + 32 \gamma_E L_{fr} \omega^2
          - 32 \gamma_E L_{qr}
          + 56 \gamma_E^2
          \bigg\}

       + \beta_0^4 C_A   \bigg\{
           16 \zeta_2
          - 4 L_{fr}^2
          + 8 L_{fr}^2 \omega
          - 4 L_{fr}^2 \omega^2
          - 4 L_{qr}
          + 4 L_{qr}^2
          + 8 \gamma_E
          - 16 \gamma_E L_{qr}
          + 16 \gamma_E^2
          \bigg\}\Bigg]\,,
   
\end{autobreak}\\ 
  \begin{autobreak}
   h^g_{d,21}(\omega) =

  \frac{1}{2 \beta_0^2 (\omega -1)^2} \beta_1 \Bigg[
        \beta_1 C_A^2   \bigg\{
          - 32 \omega
          + 16 \omega^2
          - 32 L_\omega
          \bigg\}

       + \beta_0 C_A^2 n_f   \bigg\{
          - \frac{640}{9} \omega
          + \frac{320}{9} \omega^2
          \bigg\}

       + \beta_0 C_A^3   \bigg\{
           \frac{4288}{9} \omega
          - 128 \omega \zeta_2
          - \frac{2144}{9} \omega^2
          + 64 \omega^2 \zeta_2
          \bigg\}

       + \beta_0^2 C_A n_f   \bigg\{
           \frac{4}{3}
          \bigg\}

       + \beta_0^2 C_A^2   \bigg\{
          - \frac{4}{3}
          + 32 L_{fr}
          - 64 L_{fr} \omega
          + 32 L_{fr} \omega^2
          - 32 L_{qr}
          + 48 \gamma_E
          \bigg\} \Bigg]\,,

\end{autobreak}\\ 
  \begin{autobreak}
   h^g_{d,22}(\omega) =
         \frac{1}{2 \beta_0 (\omega -1)^3} \Bigg[ C_A^2 n_f   \bigg\{
           \frac{32}{27} \omega
          \bigg\}

       + C_A^3   \bigg\{
          - \frac{176}{27} \omega
          \bigg\}\Bigg]\,,

\end{autobreak}
\end{align}
where $\gamma_E$ is the Euler-Mascheroni constant. Here, ${L}_{\omega}=\ln(1-\omega)$ with $\omega = \beta_0 a_s(\mu_R^2) \ln N_1 N_2$, $L_{qr} = \ln \big(\frac{q^2}{\mu_R^2}\big)$ and $L_{fr} = \ln \big(\frac{\mu_F^2}{\mu_R^2}\big)$.
\end{widetext}


\bibliography{main}
\bibliographystyle{apsrev4-1}
\end{document}